\documentclass{pasj00}
\draft
\usepackage{subfigure}

\begin{document}
\SetRunningHead{Author(s) in page-head}{Running Head}

\title{Direct Imaging Search for Extrasolar Planets in the Pleiades}

\author{%
	Kodai~\textsc{Yamamoto}\altaffilmark{1},
	Taro~\textsc{Matsuo}\altaffilmark{2},
	Hiroshi~\textsc{Shibai}\altaffilmark{1},
	Yoichi~\textsc{Itoh}\altaffilmark{3},
	Mihoko~\textsc{Konishi}\altaffilmark{1},
	Jun~\textsc{Sudo}\altaffilmark{1},
	Ryoko~\textsc{Tanii}\altaffilmark{4},
	Misato~\textsc{Fukagawa}\altaffilmark{1},
	Takahiro~\textsc{Sumi}\altaffilmark{1},
	Tomoyuki~\textsc{Kudo}\altaffilmark{5},	
	Jun~\textsc{Hashimoto}\altaffilmark{6},			
	Nobuhiko~\textsc{Kusakabe}\altaffilmark{6},		
	Lyu~\textsc{Abe}\altaffilmark{7},				
	Wolfgang~\textsc{Brandner}\altaffilmark{8},		
	Timothy~D.~\textsc{Brandt}\altaffilmark{9},		
	Joseph~\textsc{Carson}\altaffilmark{10},		
	Thayne~\textsc{Currie}\altaffilmark{11},			
	Sebastian~E.~\textsc{Egner}\altaffilmark{5},		
	Markus~\textsc{Feldt}\altaffilmark{8},			
	Miwa~\textsc{Goto}\altaffilmark{8},				
	Carol~\textsc{Grady}\altaffilmark{12},			
	Olivier~\textsc{Guyon}\altaffilmark{5},			
	Yutaka~\textsc{Hayano}\altaffilmark{5},			
	Masahiko~\textsc{Hayashi}\altaffilmark{13},		
	Saeko~\textsc{Hayashi}\altaffilmark{5},			
	Thomas~\textsc{Henning}\altaffilmark{8},		
	Klaus~\textsc{Hodapp}\altaffilmark{14},			
	Miki~\textsc{Ishii}\altaffilmark{5},				
	Masanori~\textsc{Iye}\altaffilmark{6},			
	Markus~\textsc{Janson}\altaffilmark{9},			
	Ryo~\textsc{Kandori}\altaffilmark{6},			
	Gillian~R.~\textsc{Knapp}\altaffilmark{9},			
	Masayuki~\textsc{Kuzuhara}\altaffilmark{6},		
	Jungmi~\textsc{Kwon}\altaffilmark{15, 6}
	Mike~\textsc{McElwain}\altaffilmark{16},			
	Shoken~\textsc{Miyama}\altaffilmark{17},		
	Jun-Ichi~\textsc{Morino}\altaffilmark{6},			
	Amaya~\textsc{Moro-Martin}\altaffilmark{18},	
	June~\textsc{Nishikawa}\altaffilmark{6},
	Tetsuo~\textsc{Nishimura}\altaffilmark{5},		
	Tae-Soo~\textsc{Pyo}\altaffilmark{5},			
	Eugene~\textsc{Serabyn}\altaffilmark{19},		
	Hiroshi~\textsc{Suto}\altaffilmark{6},			
	Ryuji~\textsc{Suzuki}\altaffilmark{20},			
	Michihiro~\textsc{Takami}\altaffilmark{21},		
	Naruhisa~\textsc{Takato}\altaffilmark{5},		
	Hiroshi~\textsc{Terada}\altaffilmark{5},			
	Christian~\textsc{Thalmann}\altaffilmark{8},		
	Daigo~\textsc{Tomono}\altaffilmark{5},			
	Edwin~L.~\textsc{Turner}\altaffilmark{10},		
	John~\textsc{Wisniewski}\altaffilmark{22},	
	Makoto~\textsc{Watanabe}\altaffilmark{23},		
	Toru~\textsc{Yamada}\altaffilmark{24},			
	Hideki~\textsc{Takami}\altaffilmark{5},			
	Tomonori~\textsc{Usuda}\altaffilmark{5},		
	and 
	Motohide~\textsc{Tamura}\altaffilmark{6}}

\altaffiltext{1}{Department of Earth and Space Science, Graduate School of Science, Osaka University,
 1-1 Machikaneyama, Toyonaka, Osaka 560-0043, Japan}
\email{yamamoto@iral.ess.sci.osaka-u.ac.jp}
\email{shibai@iral.ess.sci.osaka-u.ac.jp}
\email{konishi@iral.ess.sci.osaka-u.ac.jp}
\email{sudo@iral.ess.sci.osaka-u.ac.jp}
\email{misato@iral.ess.sci.osaka-u.ac.jp}
\email{sumi@iral.ess.sci.osaka-u.ac.jp}

\altaffiltext{2}{Department of Astronomy, Faculty of Science, Kyoto University, Kitashirakawa-Oiwake-cho, Sakyo-ku, Kyoto 606-8502, Japan}
\email{matsuo@kusastro.kyoto-u.ac.jp}

\altaffiltext{3}{Nishi-Harima Astronomical Observatory, 407-2 Nishigaichi, Sayo-cho, Sayo-gun, Hyogo 679-5313, Japan}
\email{yitoh@nhao.jp}

\altaffiltext{4}{Graduate School of Science, Kobe University, 1-1 Rokkodai, Nada, Kobe, Hyogo 657-8501, Japan}

\altaffiltext{5}{Subaru Telescope, 650 North Aohoku Place, Hilo, HI 96720, USA}	

\altaffiltext{6}{National Astronomical Observatory of Japan, 2-21-1 Osawa, Mitaka, Tokyo 181-8588, Japan}
\email{motohide.tamura@nao.jp}

\altaffiltext{7}{Laboratoire Lagrange, UMR7293, Universit$\acute{e}$ de Nice-Sophia Antipolis, CNRS, Observatoire de la C$\hat{o}$te d'Azur, 06300 Nice, France}

\altaffiltext{8}{Max Planck Institute for Astronomy, Heidelberg, Germany}

\altaffiltext{9}{Department of Astrophysical Sciences, Princeton University, NJ 08544, USA}

\altaffiltext{10}{Department of Physics and Astronomy, College of Charleston, 58 Coming St., Charleston, SC 29424, USA}

\altaffiltext{11}{Department of Astronomy and Astrophysics, University of Toronto, 27 King's College Circle, Toronto, Ontario, Canada M5S 1A1}

\altaffiltext{12}{Eureka Scientic, 2452 Delmer, Suite 100, Oakland CA 96002, USA}

\altaffiltext{13}{Department of Astronomy, The University of Tokyo, Hongo 7-3-1, Bunkyo-ku, Tokyo 113-0033, Japan}

\altaffiltext{14}{Institute for Astronomy, University of Hawaii, 640 North A'ohoku Place, Hilo, HI 96720, USA}

\altaffiltext{15}{Department of Astronomical Science, The Graduate University for Advanced Studies (SOKENDAI), 2-21-1 Osawa, Mitaka, Tokyo 181-8588, Japan}

\altaffiltext{16}{ExoPlanets and Stellar Astrophysics Laboratory, Code 667, Goddard Space Flight Center, Greenbelt, MD 20771, USA}

\altaffiltext{17}{Office of the President, Hiroshima University, 1-3-2 Kagamiyama, Higashi-Hiroshima, 739-8511, Japan}

\altaffiltext{18}{Departamento de Astrofisica, CAB (INTA-CSIC), Instituto Nacional de T$\acute{e}$cnica Aeroespacial, Torrej$\acute{o}$n de Ardoz, 28850, Madrid, Spain}

\altaffiltext{19}{Jet Propulsion Laboratory, California Institute of Technology, Pasadena, CA, USA}

\altaffiltext{20}{TMT Observatory Corporation, 1111 South Arroyo Parkway, Pasadena, CA 91105, USA}

\altaffiltext{21}{Institute of Astronomy and Astrophysics, Academia Sinica, P.O. Box 23-141, Taipei 106, Taiwan}

\altaffiltext{22}{Department of Astronomy, University of Washington, Box 351580 Seattle, WA 98195, USA}

\altaffiltext{23}{Department of Cosmosciences, Hokkaido University, Sapporo 060-0810, Japan}

\altaffiltext{24}{Astronomical Institute, Tohoku University, Aoba, Sendai 980-8578, Japan}


\maketitle

\begin{abstract}
We carried out an imaging survey for extrasolar planets around stars in the Pleiades (125~Myr, 135~pc) in the $H$ and $K_\mathrm{S}$ bands using HiCIAO combined with the adaptive optics, AO188, 
on the Subaru telescope. 
We found 13 companion candidates fainter than 14.5 mag in the $H$ band around 9 stars.
Five of these 13 were confirmed to be background stars by measurement of their proper motion. 
One was not found in the second epoch observation, and thus was not a background or companion object.
One had multi-epoch image, but the precision of its proper motion was not sufficient to conclude whether it was background object. 
Four other candidates are waiting for second epoch observations to determine their proper motion. 
Finally, the remaining 2 were confirmed to be 60~$M_\mathrm{{J}}$ brown dwarf companions orbiting around HD~23514 (G0) and HII~1348 (K5) respectively, as had been reported in previous studies. 
In our observations, the average detection limit for a point source was 20.3 mag in the $H$ band beyond \timeform{1''.5} from the central star. 
On the basis of this detection limit, we calculated the detection efficiency to be 90\% for a planet with 6 to 12 Jovian masses and a semi-major axis of 50--1000 AU. 
For this we extrapolated the distribution of planet mass and semi-major axis derived from RV observations and adopted the planet evolution model of \citet{Baraffe+2003}.
As there was no detection of a planet, we estimated the frequency of such planets to be less than 17.9\% ($2\sigma$) around one star of the Pleiades cluster.
\end{abstract}

\section{Introduction}
Understanding planet-building and their evolutionary process is one of the most challenging problems in astrophysics. 
Theoretically, there have been two main competing hypotheses regarding the formation of gas-giant planets: 
core accretion (e.g., \cite{Safronov1969, Mizuno1980, Pollack+1996}) and disk instability (e.g., \cite{Kuiper1951, Cameron1978}). 
Planet formation theories have been continuously updated or newly proposed (e.g., \cite{Inutsuka+2010}), but these two hypotheses have served as the basis for most studies. 
On the one hand, in the core accretion model, relatively small giant planets such as Jupiter and Saturn are thought to form at about 10~AU or less from a solar-type host star in several Myr (\cite{Pollack+1996, IdaLin2004}). 
On the other hand, in the disk instability model, planets of a few to 10 $M_\mathrm{{J}}$ can be created within a few 10 to 100~AU from the central star on a dynamical timescale of several thousand years (\cite{Rafikov2007, Rafikov2011, Marois+2008, Kratter+2010, Janson+2012}). 
These formation models therefore predict two populations of giant planets segregated by orbital distance, with the closer planets formed by core accretion and the outer ones by disk instability.

However, planets may experience subsequent orbital migration as a result of interaction with the parent disk either inward or even outward in the case of type III migration (\cite{MassetPapaloizou2003}). 
Furthermore, in a system with multiple planets, one can be ejected beyond the outer radius of the disk through gravitational interaction between planets or their embryos (e.g. \cite{IdaLin2004, Veras+2009, BasuVorobyov2012}).
In addition, free-floating planets might be captured at wide orbits, although such widely separated planets are likely rare (on the order of a few percent, e.g., \cite{Kouwenhoven+2010}). 
Thus, a number of mechanisms to explain the formation and evolution of planets have been theoretically explored, but it is most important to observationally determine planet frequency over a wide range of orbital distances.

Observationally, more than 830 extrasolar planets have been found to date, of which about 90\% were detected by radial velocity (RV) and transit observations (e.g. \cite{Mayor+2011, Howard+2010}). 
This rapidly growing sample allows a statistical discussion of planet frequency based on the properties of the planets and their host stars. 
However, these observing methods have a limitation: it is difficult to detect planets that are far from host stars, i.e., more than about 10~AU. 
Direct imaging, however, which is sensitive to such distant regions, 
 can provide critical and complementary information to that obtained by indirect detection methods (\cite{Marois+2008, Marois+2010, Lagrange+2010, Currie+2011, Carson+2012}). 
Given its importance and with the development of instruments and observing techniques, direct imaging has been extensively performed in recent years with large-aperture telescopes. 
\citet{Lafreniere+2007} calculated the planet frequency around a single star as less than 0.1 (for separations in the range 50--250~AU and planet masses 0.5--$13~M_\mathrm{J}$) on the basis of the Gemini observations of 85 stars. 
In \citet{Nielsen+2010}, the frequency (8.9--911~AU, $> 4~M_\mathrm{J}$) was estimated to be below 0.2, by compiling the data of 118 stars (\cite{Liu2004, Masciadri+2005, Marois+2006, Biller+2007, Lafreniere+2007}). 
Moreover, \citet{Chauvin+2010} reported VLT observations of 88 targets (10--500~AU, $>1~M_\mathrm{{J}}$) that yielded a frequency of below 0.1. 
\citet{Vigan+2012} reported the frequency of a planet around early type stars (A--F) to be $8.7^{+10.1}_{-2.8}$ (1$\sigma$).
The result of the previous direct imaging surveys for the frequency of a planet summarized in Table~\ref{tab:summary of others}.
The problem with direct imaging is that the sample size is small compared to that of indirect observations.

In these imaging studies, the targets belong to the moving groups and local associations including the $\beta$ Pictoris moving group, TW~Hya Association, Tucana-Horologium Association, and AB Doradus group (\cite{Lafreniere+2007, Chauvin+2010}). 
Because these associations are nearby ($\sim$20--100~pc) and young (several to several hundred Myr), their planets are relatively bright and should be easy to detect. 
In addition, stars in the same cluster have similar ages and distances from earth, which statistically improves the accuracy of the age and luminosity estimates, and hence the derivation of the planetary mass. 
However, the number of the group members is not large. 
For instance, such sparse moving groups have only several dozen members each, and only a dozen stars have been observed by previous studies (\cite{Chauvin+2010}). 
In contrast, open clusters usually have many more members, which can be an advantage when discussing the frequency of planets at specific ages, as well as for obtaining relatively accurate estimates of planetary masses.

\begin{table*}
  \caption{Summary of the direct imaging observations.}\label{tab:summary of others}
  \begin{center}
    \begin{tabular}{lclcclccc}
      \hline
      Author					& Sp. Type	& Target						& Age		& Distance	& Number	& \multicolumn{2}{c}{Investigated range}	& Planet  	 \\
      							& (median)	& cluster	\footnotemark[$*$]	& (Myr)		& (pc)		&			& Mass				& Separation\footnotemark[$\dagger$] 		& frequency	 \\
							&			&							& (median)	& (median)	&			& ($M_\mathrm{{J}}$)	& (AU)			& (\%)		 \\	
     \hline	
      \citet{Lafreniere+2007}	& F2--M4	& 1, 2, 3, 5,					& 10--300	& 3.2--34.9	& 85		& 0.5--13			& 50--250		& $\le$9.3	 \\
      							& (K0)		& 8, 9, 10,					& (100)		& (22)		&			&					& (sma)				&				\\
      							&			& 11, 14						&			&			&			&					&				&				\\
      \citet{Chauvin+2010} 		& B7--M8	& 2, 3, 4, 5,					& 8--100	& 10--130	& 88		& 0.5--15			& 10--500		& $<$ 10	\\
      							&			& 6, 7, 12, 14				& ()			& ()			&			&					& (pro)			&				\\
       \citet{Nielsen+2010}		& A5--M5	& 1, 2, 3, 5, 8, 	 			& 2--8800	& 3.2--77.0	& 118		& $>$ 4				& 8.9--911		& $<$ 20 		\\
      							& (K1)		& 9, 10, 11, 12,				& (160)		& (24)		&			&					& (sma)			&				\\
      							&			& 12, 13, 14					&			&			&			&					&				&				\\
      \citet{Vigan+2012}		& A0--F5	& -							& 8--400	& 19--84	& 42		& 3--14				& 5--320		& $8.7^{+10.1}_{-2.8}$\\
      							& (A3)		&							& (100)		& (50)		&			&					& (sma)			&				\\
      \hline
      \multicolumn{9}{@{}l@{}}{\hbox to 0pt{\parbox{160mm}{\footnotesize
		\par\noindent
		\footnotemark[$*$] Moving groups: 
		(1) $\alpha$ Persei; (2) AB Doradus; (3) $\beta$ Picoris; (4) Carina; (5) Carina-Near; (6) Columba; (7) $\eta$ Cha; (8) Hercules-Lyra; (9) IC2391; 
		(10) Local association; (11) Local association subgroup B4; (12) Tucana-Horologium; (13) TW Hydrae association; (14) Ursa Major.
		\par\noindent
		\footnotemark[$\dagger$] Separation:
		sma: semi-major axis; pro: projected 
      }\hss}}
    \end{tabular}
  \end{center}
\end{table*}

We therefore have started an imaging survey of planets in an open cluster, the Pleiades, in order to constrain the frequency of gas-giant planets at $>$50~AU around the member stars. 
The imaging is conducted with the near-infrared instrument HiCIAO with the AO188 adaptive optics on the Subaru telescope (\cite{Suzuki+2010, Hodapp+2008}). 
Here we report the imaging results for the first 20 surveyed stars.

\section{Target selection}
Our purpose is to detect extrasolar planets of less than 10~Jovian masses as close as possible to the central star.
Therefore, we selected the Pleiades, a nearby young star cluster observable from the northern hemisphere.
The Pleiades cluster is significantly populous and thus it provides a better probe of the planet frequency at a given age and for a given common star-formation history.
It is located at 133.5$\pm1.2$~pc (\cite{An+2007, Soderblom+2005, van Leeuwen2009}) and is 125$\pm8$~Myr~old (\cite{Stauffer+1998}).
The typical metallicity of the cluster members is similar to that of the Sun ([Fe/H] = -0.03$\pm$0.06; \cite{Gratton2000}).

One of the important criteria for choosing an open cluster is the sensitivity for detecting giant planets of $<$10~$M_\mathrm{{J}}$. 
The luminosity of a planet depends on its age and mass. 
To be consistent with previous studies, in our work, we have adopted the evolutionary model of \citet{Baraffe+2003} to predict the brightness of planets.
The $H$-band magnitudes for a planet at 125~Myr are thus estimated to be 27.9, 22.5, and 20.4~magnitudes (mag) for 1, 5, and $10~M_\mathrm{{J}}$, respectively.
The typical integration time in our observations is about 30~minutes with HiCIAO/AO188, as described later, which provides a detection limit (5$\sigma$) of 21.5 mag.
This means that it is possible to detect a planet less massive than $10~M_\mathrm{{J}}$.

We note that it has been predicted that the formation process itself is also related to the luminosity evolution of a planet.
There are two types of evolutionary models: hot start and cold start. 
Since the hot-start model assumes higher entropy for giant planets, 
it may correspond to planet formation by the collapse of a gaseous disk (\cite{Baraffe+1998, Baraffe+2002, Baraffe+2003, ChabrierBaraffe2000}),
while the cold start condition may represent core accretion process (\cite{Fortney+2005, Fortney+2008, Marley+2007}).
It has been shown that higher initial entropy causes a planet to became brighter (\cite{SpiegelBurrows2012}).
Thus, the brightness of a planet at a certain age as derived by the hot start model serves as an upper limit, while the cold start model represents a lower limit.
The model by \citet{Baraffe+2003} is a hot start model.
Based on the cold start model (\cite{SpiegelBurrows2012}), the $H$ magnitude is predicted to be 22.6 mag for a planet with $12~M_\mathrm{{J}}$, indicating that we do not have the sensitivity to detect such planets.
Since planet mass estimates are dependent on the evolutionary model that is used, we should be aware of such uncertainties.

Target stars in the Pleiades were selected on the basis of the following three criteria.
\begin{enumerate}
\item The star is brighter than 12 mag in the $R$ band.

AO imaging requires a guide star to measure and correct the atmospheric distortion in optical, so the star should be bright in $R$ to obtain diffraction-limited performance.
In the case of Subaru/AO188, the guide star needs to be located within \timeform{30''} of the target; thus, the target star itself is normally used as the AO guide star.
\item The membership probability is high.

Cluster membership for the target star is confirmed by using the following three criteria.
First, the membership probability should be higher than 80\% based on the proper motion measurements of \citet{Belikov+1998} and the target star should not be classified as a non-member by the other proper motion tests of \citet{Lodieu+2007}.
Second, if the star fails to fulfill the first sub-criterion, it needs to have a membership probability (\cite{Belikov+1998}) higher than 50\% and be determined to be a member according to \citet{Lodieu+2007}.
Third, if the star does not satisfy the above two sub-criteria, it should be classified as a Pleiades member on the basis of the proper motion and photometry of \citet{Stauffer+2007}.
\item The star has no binary companion that might exert gravitational influence on planet formation.

The target star should not be identified as a binary in literature (\cite{Bouvier+1997, RaboudMermilliod1998, Lodieu+2007}).
In addition, there should be no other bright ($<$15~mag in the $H$ band) object in the field of view (FoV) of \timeform{20''}$\times$\timeform{20''} by 2MASS observation.
\end{enumerate}
Finally, we selected 60 targets out of 455 stars in the Pleiades (\cite{Belikov+1998, Micela+1996, Pinfield+2003, RaboudMermilliod1998}).

\section{Observations}
\label{sec:observation}
\begin{table*}
  \caption{Summary of the observations.}\label{tab:summary of observations}
  \begin{center}
    \begin{tabular}{lcllcccrcr}
      \hline
      Name				& Sp. Type	& Date	& Obs. mode/		& $H$/$K_\mathrm{S}$\footnotemark[$\S$]		& R		& T$_\mathrm{exp}$	& N$_\mathrm{exp}$	& T$_\mathrm{total}$ & Ang. $_\mathrm{FoV}$\\
      						& 			&		& Filter				& (mag)										& (mag)	& (sec)		&				& (min) 	& (degree)\\
     \hline
      BD +22 574			& F8\footnotemark[$*$]	& 2009-10-31	& ADI / $H$		& 8.854					& 10.02	& 10		& 207			& 34.5 	& 116.9 \\
      HD 23912			& F3V\footnotemark[$*$]	& 2009-10-31	& ADI / $H$		& 8.097					& 8.88	& 10		& 30			& 5 		& 4.1\\
      						& 						& 2010-01-23	& ADI / $H$		& 						& 		& 10		& 175			& 29.2 	& 72.8\\
      						& 						& 2011-01-27	& DI / 	$H$		& 						& 		& 10		& 30			& 5 		& -\\
      V1171 Tau			& G8\footnotemark[$\dagger$]	& 2009-11-01	& ADI / $H$		& 9.270			& 10.58	& 10		& 30			& 5 		& 28.1\\
      						&						& 2012-12-31	& DI / $H$		&						&		& 30		& 15			& 7.5	& -\\
      HII 2462				& G2\footnotemark[$\dagger$]	& 2009-12-22	& ADI / $H$		& 9.699			& 10.87	& 10		& 60			& 10 	& 52.8\\
      HD 23863			& A7V\footnotemark[$*$]	& 2009-12-23 	& ADI / $H$		& 7.599					& 7.98	& 10		& 93			& 15.5 	& 46.3\\
      HD 282954			& G0\footnotemark[$\dagger$]	& 2010-01-24 	& ADI / $H$		& 8.851			& 9.98	& 10		& 223			& 37.2 	& 90.9\\
      						&						& 2012-09-12	& DI / $H$		&						&		& 2.5		& 36			& 1.5	& - \\
      HD 23514			& G0\footnotemark[$*$]	& 2010-12-01 	& ADI / $H$		& 8.291					& 8.96	& 10		& 204			& 34 	& 147.6\\
      HD 23247			& F3V\footnotemark[$*$]	& 2011-01-27 	& ADI / $H$		& 7.811					& 8.85	& 10		& 83			& 13.8 	& 79.7\\
      						& 						& 2011-12-23 	& ADI / $H$		&						&		& 10		& 65			& 10.8 	& -\\
      V855 Tau			& F8\footnotemark[$\dagger$]	& 2011-01-28 	& ADI / $H$		& 8.337				& 9.37	& 10		& 160			& 26.7 	& 114.8\\
      						& 						& 2012-01-01 	& DI / 	$H$		& 						&		& 10		& 270			& 45 	& -\\
      HD 24132			& F2V\footnotemark[$*$]	& 2011-01-29 	& ADI / $H$		& 7.930					& 8.59	& 10		& 134			& 22.3 	& 107.9\\
      HD 23061			& F5V\footnotemark[$*$]	& 2011-01-30 	& ADI / $H$		& 8.325					& 9.28	& 10		& 149			& 24.8 	& 103.5\\
      TYC 1800-2144-1		& G0V\footnotemark[$\dagger$]	& 2011-01-31 	& ADI / $K_\mathrm{S}$	& 8.868	& 10.37	& 10		& 58			& 9.7 	& 72.5\\
      HII 1348				& K5\footnotemark[$\dagger$]	& 2011-12-23 	& ADI / $H$		& 9.831			& 11.92	& 10		& 141			& 23.5 	& 90.4\\
      Melotte 22 SSHJ G214	& G2\footnotemark[$\dagger$]	& 2011-12-23 	& ADI / $H$		& 9.634			& 11.17	& 10		& 180			& 30 	& 59.1\\
      BD +23 514			& G5\footnotemark[$\dagger$]	& 2011-12-24 	& ADI / $H$		& 9.528			& 10.90	& 10		& 121			& 20.2 	& 97.3\\
      Melotte 22 SSHJ G213	& G2\footnotemark[$\dagger$]	& 2011-12-24 	& ADI / $H$		& 9.543			& 10.91	& 5			& 410			& 34.2 	& 31.4\\
      Melotte 22 SSHJ G221	& G2IV\footnotemark[$\ddagger$]	& 2011-12-25 & ADI / $H$	& 9.311				& 10.76	& 10		& 270			& 45 	& 41.9\\
      V1054 Tau			& --						& 2011-12-30 	& ADI / $H$		& 9.921					& 11.35	& 10		& 150			& 25.8 	& 105.2\\
      V1174 Tau			& --						& 2011-12-30 	& ADI / $H$		& 10.197				& 11.61	& 10		& 170			& 28.3 	& 21.3\\
      Melotte 22 SSHJ K101	& --						& 2011-12-31 	& ADI / $H$		& 9.959					& 11.69	& 10		& 80			& 13.3 	& 58.4\\
      \hline
      \multicolumn{10}{@{}l@{}}{\hbox to 0pt{\parbox{160mm}{\footnotesize
		\par\noindent
      		DI; direct imaging. ADI; angular differential imaging.
		$T_\mathrm{exp}$; integration time of each exposure. 
		$N_\mathrm{exp}$; total number of exposures. 	
		$T_\mathrm{total}$; total exposure time. 
		Ang. $_\mathrm{FoV}$; rotation angle of field of view during observation.
		\par\noindent
		\footnotemark[$*$] \cite{Wright+2003}
		\par\noindent
		\footnotemark[$\dagger$] \cite{Skiff2010}
		\par\noindent
		\footnotemark[$\ddagger$] \cite{Belikov+2002}
		\par\noindent
		\footnotemark[$\S$] Hmag; Cutri et al. 2003, Rmag; Zacharias et al. 2005
      }\hss}}
    \end{tabular}
  \end{center}
\end{table*}
Twenty of the 60 selected target stars were observed between October 2009 and January 2012 (Table~\ref{tab:summary of observations}). 
The imaging observations were carried out as part of the Strategic Explorations of Exoplanets and Disks with Subaru (SEEDS, \cite{Tamura2009}) by using HiCIAO, which is a high-contrast instrument installed on the Subaru telescope (\cite{Suzuki+2010, Hodapp+2008}). 
HiCIAO has a 2048$\times$2048 HgCdTe/HAWAII 2 detector array and its pixel scale is 9.5 mas/pixel; thus, the FoV is $\sim$ \timeform{20''}$\times$\timeform{20''}. 
The targets were observed either with the $H$ or $K_\mathrm{S}$ filter. 
The coronagraphic masks were not used. 

To obtain the high contrast needed to observe within the close vicinity of a host star, HiCIAO was used in combination with AO188 (\cite{Hayano+2010}). 
By using AO, a FWHM of 6--10 pixels (\timeform{0''.05}--\timeform{0''.10}) was achieved for a point source. 
In addition, angular differential imaging (ADI; \cite{Marois+2006}) was implemented. 
ADI is an imaging method that allows the rotation of the FoV with time but fixes the detector plane relative to the pupil plane by using an image-rotator. 
As a result, this method can effectively reduce quasi-static noise including the halo of the star and speckles produced by the telescope, because the noise pattern is fixed on the detector. 
The key to obtaining effective noise reduction is a large field rotation; 
therefore, the imaging was performed to cover the period of transit of the target stars over the meridian, giving a rotation angle of 25--150 degrees. 
Additionally, the target star was placed at the center of the FoV to provide a wide area for the planet search.

Our observational procedure consisted of three steps. 
First, 5--10 unsaturated frames were taken as a reference for the point-spread-function (PSF) of the central star with 1.5 to 2.5~s exposure time to avoid saturation. 
Second, the ADI observations were performed over an integration of 5 or 10 s in the individual frames to obtain high sensitivity, but with no smearing caused by the field rotation. 
The central star was saturated at the peak by this integration time, and the saturated area had a radius of 3--6 pixels. 
Third, several unsaturated frames were retaken. 
Table~\ref{tab:summary of observations} summarizes the information on the observed stars, observing mode, filters, and exposure times of saturated images.

If sources were detected around a target star, they were considered to be candidate companions (CCs). 
For HD~23247, the bright ($H<$ 14.5) companion candidate was detected at \timeform{3''.7} from the central star.
However, we discuss only CCs fainter than 14.5 mag which corresponds to about 100 $M_\mathrm{{J}}$ (brown dwarf mass), in the subsequent part of this paper since our focus is not on the stellar regime.
The relative positions of CCs against the target star were measured in the follow-up observations for HD~23912 and V855~Tau to determine whether they were co-moving. 
In the follow-up observations, the direct imaging (DI) mode without field rotation was employed since the CCs have wide angular separation (more than about 3~arcsec). 
V1171~Tau, BD+22~574, and HD~282954 have been observed with a different camera, Subaru/CIAO, in 2005, and the same CCs were detected (\cite{Itoh+2011}). 
Thus, our HiCIAO observations gave the proper motion measurements combined with the CIAO results. 
HD~23912 was observed three times (in October 2009, January 2010, and January 2011). 
Since the field rotation by ADI was too small ($\sim10$ degree) for the first imaging in October 2009, it was revisited in January 2010.

\section{Data reduction}
\label{sec:data reduction}
The first step of the image processing was to remove the striped pattern caused by fluctuations in the bias levels in the individual raw images. 
The stripes consist of two components: 
32 horizontal stripes each with a height of 64 pixels, and thin vertical stripes, each 2048 pixels high, randomly distributed over the image. 
These patterns vary with time and are independent among images. 
We created the striped pattern for the whole FoV by using the sky region in each frame, and subtracting it from the raw frame, 
a process corresponding to sky subtraction. 
Next, the bad pixels and their clusters were corrected by subtracting the de-striped dark image. 
Then, we performed flat-fielding by using the dome-flats. 
Bad pixels randomly occurring in arbitrary pixel positions were interpolated from the surrounding pixels. 
These calibrations were carried out by using our own reduction tool for HiCIAO data.

The image processing that follows (described below) was performed with IRAF\footnote{IRAF is distributed by the National Optical Astronomy Observatories, 
which are operated by the Association of Universities for Research in Astronomy, Inc., under cooperative agreement with the National Science Foundation.}. 
Sub-pixel shifts cannot be avoided during the process of distortion correction and ADI reductions. 
They require the interpolation of adjacent pixels, which causes the smearing of pixel values. 
As a result, the noise level is reduced. 
Moreover, the amount of sub-pixel shift was different for each frame, and we confirmed that the degree of noise reduction could vary among multiple images. 
Such a non-uniform process, as well as artificial noise reduction, may affect our discussion of detection limits. 
Thus, before applying the distortion correction, all images were smoothed with a 2-D Gaussian filter with an FWHM of 3 pixels to obtain the same level of noise reduction for all pixels and images. 
The distortion was measured by comparison of images of the globular clusters (M5 and M15) with HiCIAO and HST/STIS \citep{van der Marel+2002}.
The distortion was then corrected to obtain a pixel scale of 9.500 $\pm$ 0.005 mas/pixel. 
The precision of the distortion correction is as described below.

Next, in preparation for the ADI reductions, the stellar position was measured and matched to the image center for all the frames. 
The target stars were saturated in areas of 3--6 pixels in radius. 
For removal of the stellar halo by ADI reductions, we adopted the centroid position of the halo measured from 10 to 50 pixels in radius as the stellar position. 
For the ADI processing, we followed the standard ADI procedures described by \citet{Marois+2006}. 
First, a reference image was created by calculating the median at each pixel position using all images. 
Second, the reference was subtracted from the individual frames. 
The resultant image was de-rotated to align the field so that north was on the top. 
Finally, the de-rotated images were median-combined with 5$\sigma$ clipping to obtain enough sensitivity to detect planetary-mass objects.

\begin{figure}
  \begin{center}
   \FigureFile(80mm,80mm){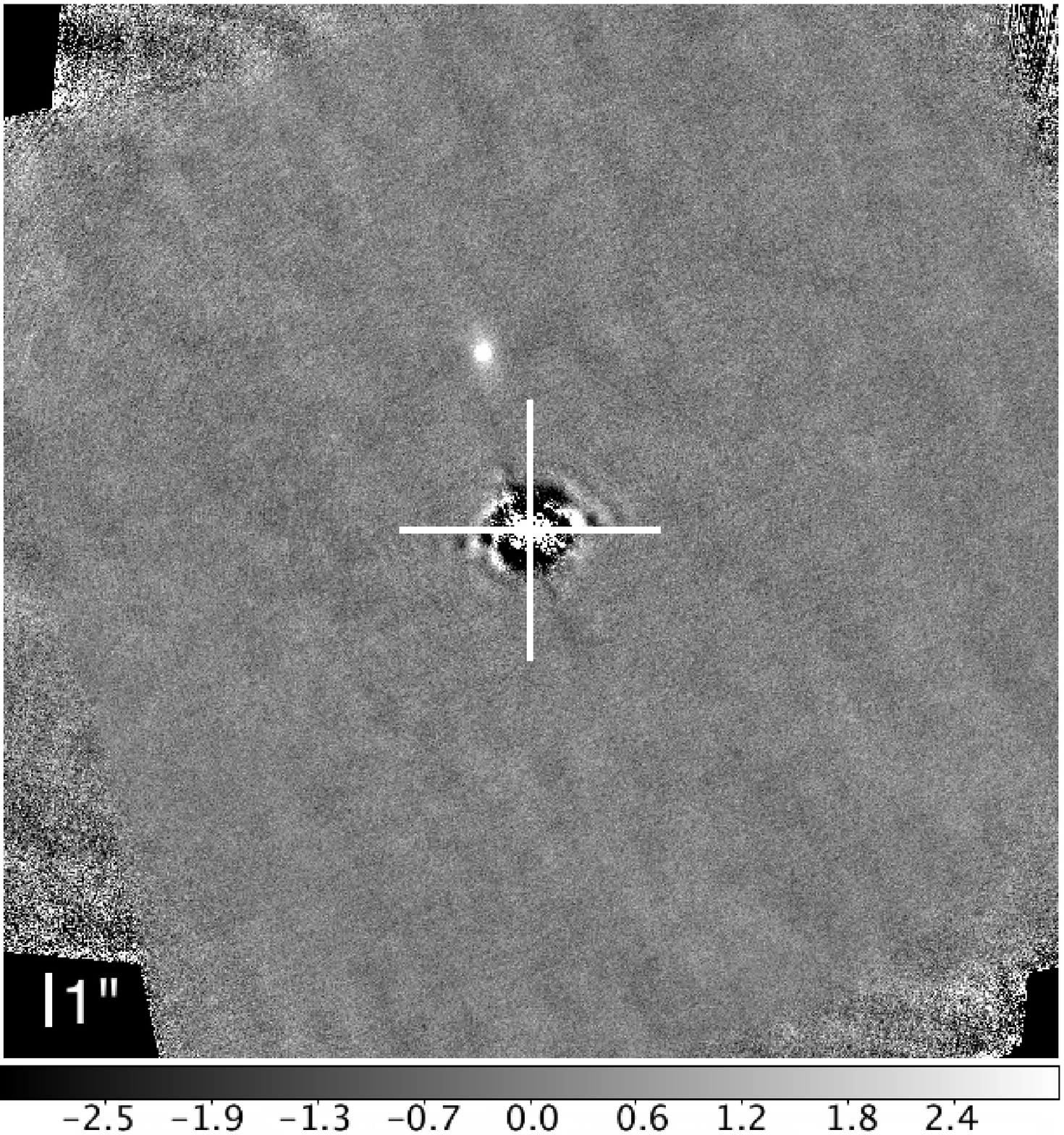}
   \FigureFile(80mm,80mm){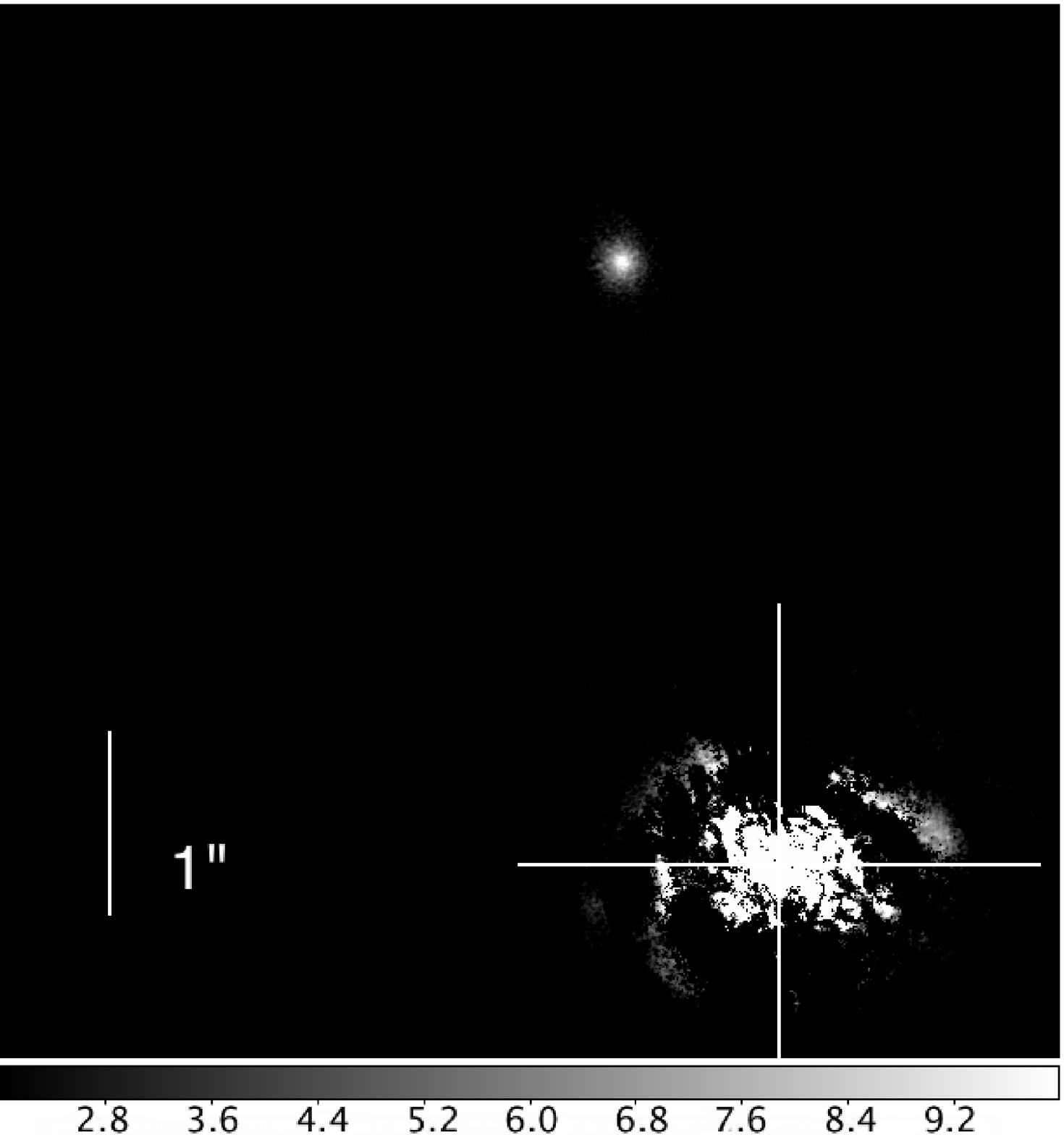}
  \end{center}
  \caption{
  The result of ADI reductions for HD~23912. 
  The image was obtained in the $H$ band. 
  The white cross indicates the position of the central star. 
  A point source was detected at an angular separation of \timeform{3''38}. 
  North is on the top and east is to the left. 
  {\it Left panel}: The field of view is \timeform{19''.5}$\times$\timeform{19''.5}. 
  The pixel value range is -0.4 to +0.6 ADU. 
  The four corners cannot be discussed since these regions are outside the FoV in many frames. 
  {\it Right panel}: The zoom-in image of the companion candidate. 
  The field of view is \timeform{5''.8}$\times$\timeform{5''.8}. 
  The pixel value range is +0.0 to +5.0 ADU. 
    }\label{fig:image of HD23912}
\end{figure}

An example of the final reduced image is presented in Figure~\ref{fig:image of HD23912}. 
The image was obtained using data from HD~23912 in the $H$ band taken in January 2010. 
The rotation angle of the FoV was 73 degrees and the total integration time was 29.2 minutes. 
At the center of the left image, the residual pattern of the subtraction of the stellar halo can be seen. 
A point source is detected at \timeform{3''.388} $\pm$ \timeform{0''.028} from the star with a position angle (P.A.) of \timeform{14D.92} $\pm$ \timeform{0D.48}. 
During the ADI, more images are taken at similar P.A. of the field when the rotation is slow. 
The emission from point sources in such images cannot be completely eliminated in the reference image, and consequently, self-subtraction occurs in the faint outskirts of the point source. 
This is consistent with the sculpting along the azimuthal direction. 
Images of all the target stars are shown in the Appendix.

When a CC was detected, the relative position between the CC and the central star was measured. 
The centroid position of the CC was determined using an aperture with radius of 1 FWHM. 
A position measurement was performed in each frame or combined frame depending on the brightness of the CC. 
In order to determine the position of the central star in saturated images, we first, 
in unsaturated images, determine the offset between the center derived by Gaussian fitting and a centroiding algorithm with a mask equal in size to the saturated area in saturated images. 
Assuming the same offset holds true in the saturated case, we correct the measurement derived by this masked centroiding algorithm accordingly.
The uncertainty of the position measurement was checked by the deviation from the rotation center of the field by ADI. 
The relative positions were measured in each (combined) image, and the rotation center was defined as the center of the fitted circular orbit for the CC in multiple images without de-rotation. 
Moreover, the deviation between the position of the CC and its fitted circular path due to the ADI observation was below 0.7 pixels.
This deviation encompasses possible distortions left even after distortion correction (as the shape would not be perfectly circular), 
thus showing that as far as any measurable effect exists it is small.
The results of the astrometry measurements are summarized in Table~\ref{tab:companion candidate}.

The magnitude of the CCs was estimated by the target star as the flux calibrator. 
The magnitudes of both for the central stars and the CCs were measured by aperture photometry.
For the central star, photometry was performed by using the unsaturated frames taken before and after the ADI observations as mentioned in section \ref{sec:observation}.
The background level was estimated as the centroid of the pixel-values histogram in an annulus with a radius of 50 pixels and the width of 20 pixels. 
The aperture size varied from 2 to 40 pixels in radius, and the converged magnitude, at a radius of about 20 pixels depending on the targets, was taken to be its magnitudes. 
By comparing this instrumental magnitude with the 2MASS measurement under the assumption that the star was not variable, the conversion from ADU to magnitude was obtained. 
The photometry for the CCs was performed with the same aperture size as that for the central star.
The flux loss by the image processing including the ADI reductions was $\sim 5$\%, 
estimated by embedding an artificial point source at radially equally spaced angles and distances (interval of \timeform{1''}) in the raw image and applying the same reduction procedures. 
The flux loss was independent of the separation beyond \timeform{1''}.
The photometry result obtained for the CCs was corrected for this flux loss. 
Finally, the magnitude of the CCs was calculated using the conversion from ADU to the magnitude derived from the photometry of the central star. 
To improve the signal-to-noise ratio (S/N), the photometry for a CC was performed with images in which 20--40 frames were combined and the results were averaged. 
The $H$ magnitudes for the CCs are shown in Table~\ref{tab:companion candidate}.

\begin{table*}
  \caption{Astrometry and photometry of companion candidates.}\label{tab:companion candidate}
  \begin{center}
    \begin{tabular}{lcllllc}
      \hline
      Name			&	Separation Angle		&	P.A.					&	$H$ 						&	Mass\footnotemark[$*$]	& UT Date 			& Status\\
      					&	 (\timeform{''})		&	($^{\circ}$ E of N)	&	(mag)						&	 ($M_\mathrm{{J}}$)		&					&\\
      \hline
      V1171 Tau CC1	& 12.770 $\pm$ 0.025	& 135.50 $\pm$ 0.40		& 18.3  \footnotemark[$\dagger$]	& -							& 2005-11-17\footnotemark[$\ddagger$]	& -\\
      					& 12.629 $\pm$ 0.028	& 134.75 $\pm$ 0.10		& 17.8 $\pm$ 0.1					& 22						& 2009-11-01 							& -\\
					& 12.603 $\pm$ 0.031	& 134.08 $\pm$ 0.21		& 17.8 $\pm$ 0.3					& 22						& 2012-12-31							& B\\
      V1171 Tau CC2	& 12.880 $\pm$ 0.027	& 136.77 $\pm$ 0.40		& 18.3  \footnotemark[$\dagger$]	& -							& 2005-11-17\footnotemark[$\ddagger$]	& -\\
      					& 12.744 $\pm$ 0.020	& 136.15 $\pm$ 0.10		& 18.5 $\pm$ 0.6					& 19						& 2009-11-01							& -\\
					& 12.628 $\pm$ 0.031	& 135.51 $\pm$ 0.21		& 18.5 $\pm$ 0.6					& 19						& 2012-12-31							& B\\
      HD 23912 CC1	& 3.388 $\pm$ 0.028		& 14.92 $\pm$ 0.48		& 17.4 $\pm$ 0.1					& 26						& 2010-01-23							& -\\
      					& 3.435 $\pm$ 0.008		& 14.52 $\pm$ 0.28		& 17.2 $\pm$ 0.2					& 28						& 2011-01-27							& B\\
      BD +22 574 CC1	& 3.405 $\pm$ 0.025		& 95.70 $\pm$ 0.20		& -\footnotemark[$\S$]			& -							& 2005-11-17							& -\\
      					& 3.288 $\pm$ 0.033		& 92.57 $\pm$ 0.20		& 19.2 $\pm$ 0.2					& 13						& 2009-10-31							& Probably B.\\
      BD +22 574 CC2	& 8.440 $\pm$ 0.030		& 51.82 $\pm$ 0.10		& 18.6 \footnotemark[$\|$]		& 14						& 2005-11-17							& -\\
      					& 8.501 $\pm$ 0.033		& 50.01 $\pm$ 0.10		& 17.4 $\pm$ 0.2					& 26						& 2009-10-31							& U\\
      HD 282954 CC1	& 9.006 $\pm$ 0.030		& 103.82 $\pm$ 0.50		& 16.4	\footnotemark[$\|$]		& 33						& 2005-11-17							& -\\
      					& 9.031 $\pm$ 0.014		& 103.23 $\pm$ 0.18		& 14.6 $\pm$ 0.1					& 87						& 2010-01-23							& -\\
					& 8.943 $\pm$ 0.014		& 103.28 $\pm$ 0.20		& 14.4 $\pm$ 0.2					& 99						& 2012-09-12							& B\\
      V855 Tau CC1	& 8.05 $\pm$ 0.03		& 19.46 $\pm$ 0.21		& 17.2 $\pm$ 0.4					& 27						& 2011-01-28							& -\\
      					& -						& -						& -								& -							& 2012-01-01\footnotemark[$\#$]			& ?\\
      HD23514 CC1	& 2.64	$\pm$ 0.02		& 228.7 $\pm$ 1.0		& - 								& - 							& 2006-12-10\footnotemark[$**$]			& -\\
      					& 2.64	$\pm$ 0.01		& 227.8 $\pm$ 0.3		& - 								& - 							& 2007-10-25\footnotemark[$**$]			& -\\
      					& 2.62	$\pm$ 0.04		& 227.2 $\pm$ 0.5		& - 								& - 							& 2008-11-04\footnotemark[$**$]			& -\\
      					& 2.642	$\pm$ 0.040		& 227.51 $\pm$ 0.04		& 15.61 $\pm$ 0.08				& 52						& 2009-11-01\footnotemark[$**$]			& -\\
      					& 2.644	$\pm$ 0.002		& 227.48 $\pm$ 0.05		& 15.39 $\pm$ 0.06				& 58						& 2010-10-30\footnotemark[$**$]			& -\\
      					& 2.646	$\pm$ 0.033		& 227.59 $\pm$ 0.72		& 15.37 $\pm$ 0.05				& 58						& 2010-12-01							& C\\
      HII 1348 CC1		& 1.09	$\pm$ 0.02		& 347.9	$\pm$ 0.7		& -								& -							& 1996-09-25 - 10-01\footnotemark[$\dagger\dagger$] & -\\
					& 1.097	$\pm$ 0.005		& 346.8 $\pm$ 0.2		& 15.30 $\pm$ 0.09				& 60						& 2004-10-03\footnotemark[$\ddagger\ddagger$] & -\\
					& 1.12 $\pm$ 0.02		& 346.8 $\pm$ 0.6 		& -								& -							& 2005-11-21\footnotemark[$\S\S$] 		& -\\
					& 1.12 $\pm$ 0.03		& 346.1 $\pm$ 0.9 		& 15.7 $\pm$ 0.4					& 48						& 2011-12-23							& C\\
      V1054 Tau CC1	& 7.110 $\pm$ 0.014		& 110.29 $\pm$ 0.11		& 18.1 $\pm$ 0.4					& 20						& 2011-12-30							& N\\
      V1054 Tau CC2	& 7.361 $\pm$ 0.028		& 76.48 $\pm$ 0.22		& 15.97 $\pm$ 0.09				& 44						& 2011-12-30							& N\\
      V1174 Tau CC1	& 6.473 $\pm$ 0.033		& 63.68 $\pm$ 0.28		& 18.0 $\pm$ 0.4					& 21						& 2011-12-30							& N\\
      V1174 Tau CC2	& 9.24 $\pm$ 0.03		& 37.4 $\pm$ 0.2			& 18.5 $\pm$ 0.3					& 17						& 2011-12-30							& N\\
      \hline
      \multicolumn{7}{@{}l@{}}{\hbox to 0pt{\parbox{160mm}{\footnotesize
		\par\noindent
      		Much brighter companion candidate was detected within \timeform{3''.7} arcsec for HD~23247.
		However, only candidates less massive than the brown-dwarf mass ($\sim$ 100 $M_\mathrm{{J}}$) are discussed in this paper.
		\par\noindent
		Status sign; U presents "undefined" due to the uncertainty of the proper motion measurement. 
		B presents the background object. 
		C presents the co-moving object. 
		N presents that the proper motion has not been measured yet.
		\par\noindent
		\footnotemark[$*$] The masses are linearly interpolated by reference to \citet{Baraffe+2003}.
		\footnotemark[$\dagger$] It was impossible to measure the individual brightness of CC1 and CC2, because they were not spatially separated not well enough for aperture photometry. 
		In addition, the error was difficult to determine due to the fluctuated PSF because of the poor seeing.
		\footnotemark[$\ddagger$] Subaru/CIAO, Subaru/IRCS \citep{Itoh+2011}.
		\footnotemark[$\S$] It was impossible to measure the brightness of the CC due to the stellar halo.
		\footnotemark[$\|$] $K$ magnitude. It was impossible to estimate the error due to the variation of PSF because of the inclement weather.
		\footnotemark[$\#$] The companion candidates were not found in the field of view. 
		\footnotemark[$**$] Keck/NIRC2 \citep{Rodriguez+2012}.
		\footnotemark[$\dagger\dagger$] CFHT/PUEO \citep{Bouvier+1997}.
		\footnotemark[$\ddagger\ddagger$] Palomar Hale telescope/PHARO \citep{Geissler+2012}.
		\footnotemark[$\S\S$] Keck/OSIRIS \citep{Geissler+2012}
      }\hss}}
    \end{tabular}
  \end{center}
\end{table*}

\section{Results}
\subsection{Detection limits of our observations}
\label{sub:obs result}
\begin{figure}
  \begin{center}
    \FigureFile(80mm,80mm){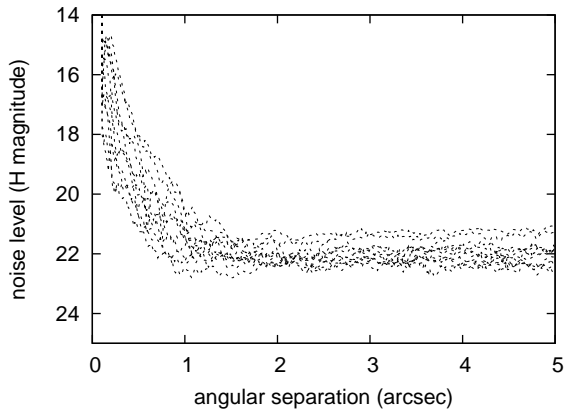}
  \end{center}
  \caption{
  Noise level (1$\sigma$) as a function of angular separation. 
  Dotted lines indicate individual observations obtained from October 2009 to January 2012. 
  The total integration time in each observation is in the range of 5--45 minutes. 
    }\label{fig:detection limit of HD23912}
\end{figure}

The detection limit of our observations is defined by a signal-to-noise ratio (S/N) of 5. 
The noise was determined by the standard deviation of the background level in the azimuthal direction measured at the same distance from the target star. 
The background level was obtained with an aperture size of approximately 2 $\times$ FWHM on the median-combined image by ADI reductions. 
The relation between the standard deviation (S/N=1) and angular separation from the central star is plotted in Figure~\ref{fig:detection limit of HD23912}. 
The median of the detection limits for all ADI observations becomes constant at 20.8 mag for S/N of 3 and 20.3 mag for S/N of 5 in the region beyond \timeform{1''.5} from the central star.
Inside $\sim$\timeform{1''.5}, the detection limit is determined by the subtraction residual of the stellar halo.
It is 17.7 mag and 19.7 mag for separations of \timeform{0''.5} and \timeform{1''.0}, respectively.

We note that there are other ways to achieve better suppression of the stellar halo than the classical ADI reductions, 
such as Locally Optimized Combination of Images (LOCI: \cite{Lafreniere+2007}). 
The LOCI algorithm considers spatial correlations of the stellar halo and speckle noise with reference images. 
However, our primary focus in this paper is on the relatively distant region from the star (more than about 100~AU) where uncorrelated, 
random noise is dominant and classical ADI is more effective than LOCI.
The results of standard ADI reductions are thus discussed in this work.

\subsection{Astrometry and photometry of companion candidates}
Among 13 companion candidates, a CC for HD~23912 is detected in our follow-up imaging with HiCIAO while the CC of V855 Tau is not found in the follow-up.
Another 7 CCs around 5 stars (BD+22~574, V1171~Tau, HD~282954, HD~23514, and HII~1348) were observed with Subaru/CIAO, Subaru/IRCS,  Keck/NIRC2,  CFHT/PUEO, Palomar Hale telescope/PHARO, Keck/OSIRIS at the previous epochs (\cite{Itoh+2011, Rodriguez+2012, Bouvier+1997, Geissler+2012}).
The relative distances to the central stars for these CCs are shown in Figure~\ref{fig:CCs}.
The remaining 4 CCs, 2 for V1054~Tau and V1174~Tau respectively, are waiting for the second epoch observations for proper motion measurements.

\begin{figure*}
  \begin{center}
  \subfigure[V1171 Tau CC1]{
    \FigureFile(80mm,80mm){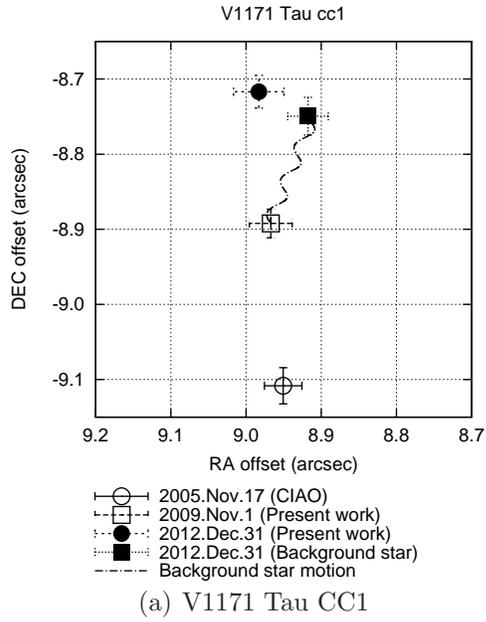}
    \label{fig:V1171TauB}}
  \subfigure[V1171 Tau CC2]{
    \FigureFile(80mm,80mm){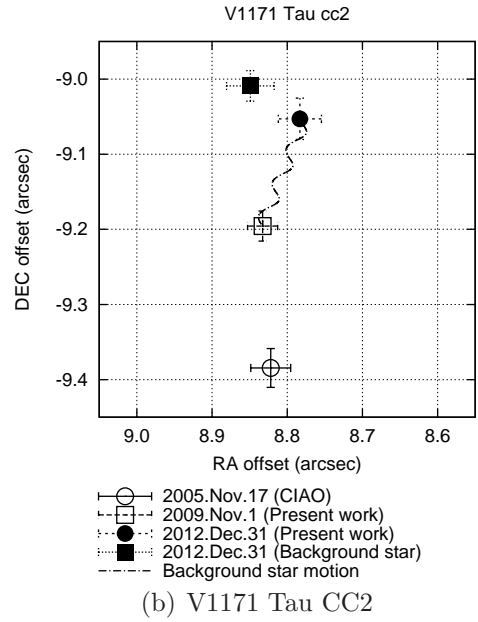}
    \label{fig:V1171TauC}}
  \end{center}
  \begin{center}
  \subfigure[HD 23912 CC1]{
    \FigureFile(80mm,80mm){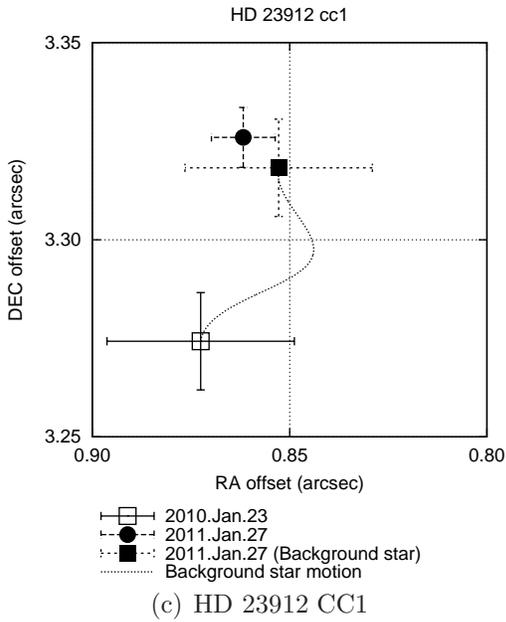}
    \label{fig:HD23912B}}
  \subfigure[BD+22 574 CC1]{
    \FigureFile(80mm,80mm){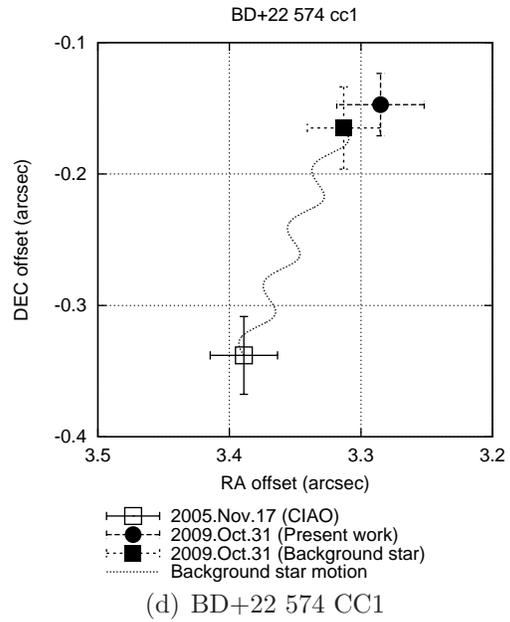}
    \label{fig:BD+22574B}}
  \end{center}
  \caption{
  	{\it Top left panel:} V1171 Tau CC1.  
  	{\it Top right panel:} V1171 Tau CC2.
  	{\it Lower left panel:} HD 23912 CC1.
  	{\it Lower right panel:} BD +22 574 CC1.
	}\label{fig:CCs}
\end{figure*}

\addtocounter{figure}{-1}
\begin{figure*}
\addtocounter{subfigure}{4}
  \begin{center}
  \subfigure[BD+22 574 CC2]{
     \FigureFile(80mm,80mm){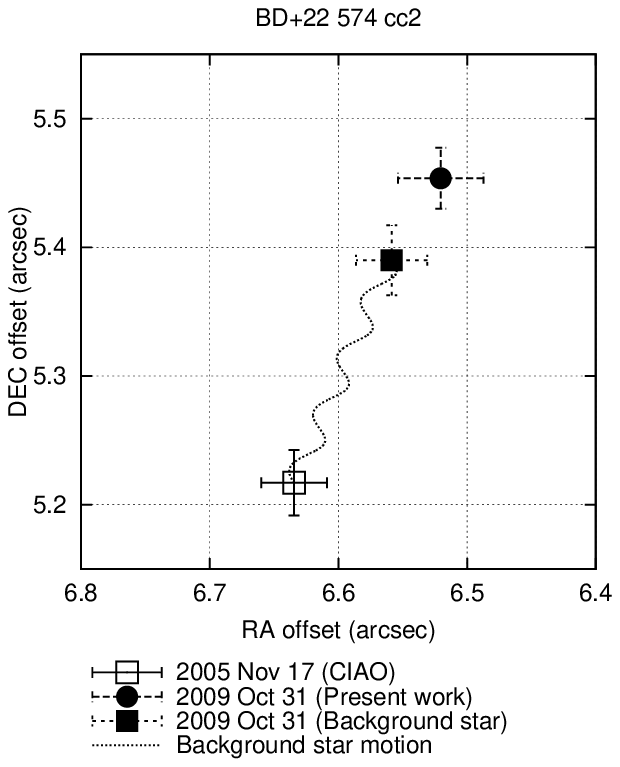}
     \label{fig:BD+22574C}}
  \subfigure[HD 282954 CC1]{
    \FigureFile(80mm,80mm){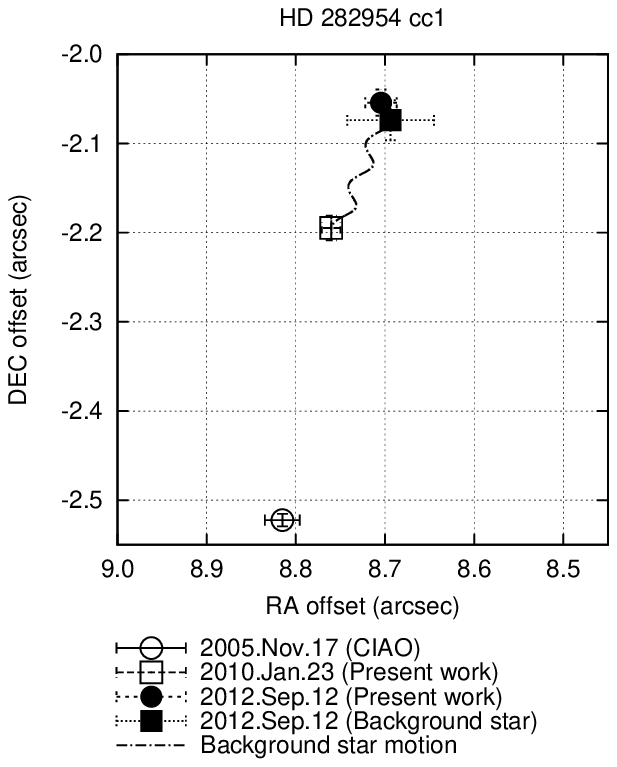}
    \label{fig:HD282954B}}
  \end{center}
  \begin{center}
  \subfigure[HD 23514 CC1]{
    \FigureFile(80mm,104mm){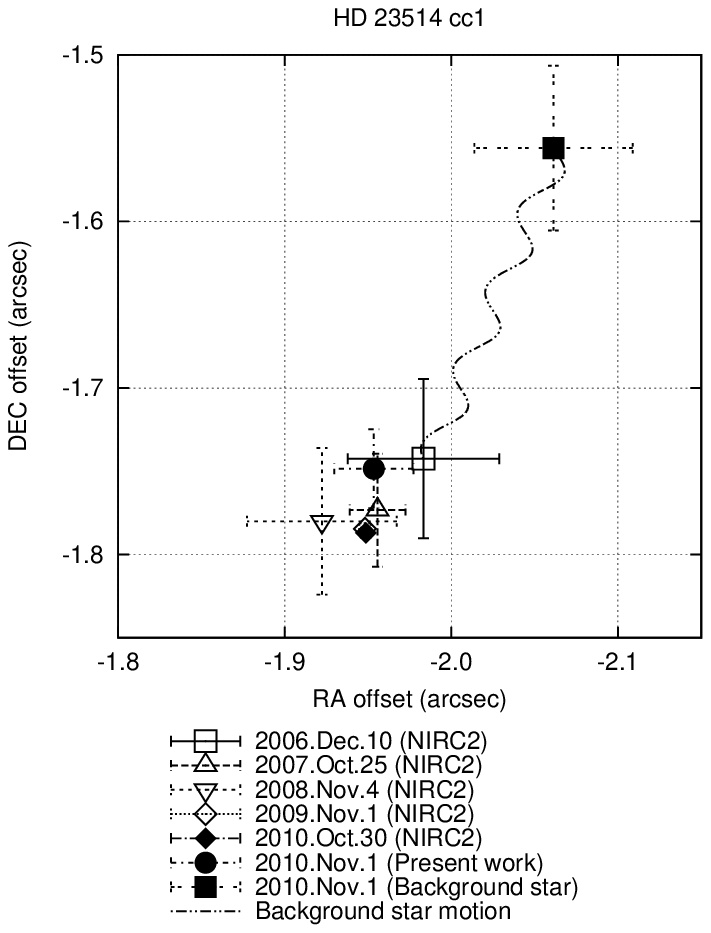}
    \label{fig:HD23514B}}
  \subfigure[HII 1348 CC1]{
    \FigureFile(77mm,86mm){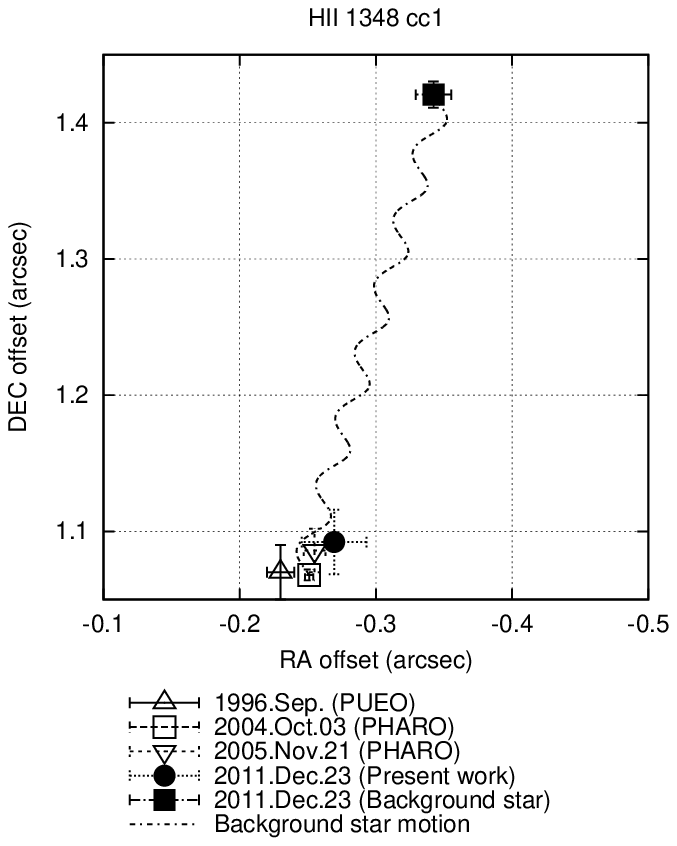}
    \label{fig:HII1348B}}
  \end{center}
  \caption{
  {\it Continued.}
		\par\noindent
  		{\it Top left panel:} BD +22 574 CC2.
		{\it Top right panel:} HD 282954 CC1.
  		{\it Lower left panel:} HD 23514 CC1.		
		{\it Lower right panel:} HII 1348 CC1.
		}\label{fig:CCs2}
\end{figure*}

\subsubsection{HD~23514, and HII~1348}
HD~23514 and HII~1348 have a co-moving object respectively, which is most likely a companion gravitationally bound to it (Figure~\ref{fig:HD23514B}, and Figure~\ref{fig:HII1348B}). 
The companion objects were first identified by the previous astrometry by \citet{Rodriguez+2012} for HD~23514, and by \citet{Geissler+2012} for HII~1348.
The $H$ magnitudes for the companion were measured to be 15.39 $\pm$ 0.06 mag in October 2010 for HD~23514, and 15.30 $\pm$ 0.09 mag in October 2004 for HII~1348, 
and their masses are estimated as 60~$M_\mathrm{{J}}$, which is in the brown dwarf regime.
 
\citet{Rodriguez+2012} measured the separation and the P.A. of HD~23514 as \timeform{2''.642} $\pm$ \timeform{0''.003} and \timeform{227D.51} $\pm$ \timeform{0D.04} in November 2009, 
and in October 2010 they were \timeform{2''.644} $\pm$ \timeform{0''.002} and \timeform{227D.48} $\pm$ \timeform{0D.05}, respectively. 
In our observation in December 2010, the separation was \timeform{2''.646} $\pm$ \timeform{0''.033} and the P.A. was \timeform{227D.6} $\pm$ \timeform{0D.7}.
The $H$ magnitude of the CC in December 2010 was 15.37 $\pm$ 0.05.
Our measurements are therefore consistent with those of \citet{Rodriguez+2012}.

For the companion of HII~1348, \citet{Geissler+2012} measured the separation and the P.A. as \timeform{1''.097} $\pm$ \timeform{0''.005} and \timeform{346D.8} $\pm$ \timeform{0D.2} in October 2004,
and in November 2005 they were \timeform{1''.12} $\pm$ \timeform{0''.02} and \timeform{346D.8} $\pm$ \timeform{0D.6}, respectively.
In our observation in December 2011, the separation was \timeform{1''.12} $\pm$ \timeform{0''.03} and the P.A. was \timeform{346D.1} $\pm$ \timeform{0D.9}.
The $H$ magnitude of the CC in December 2011 was 15.7 $\pm$ 0.4.
Our measurements are therefore consistent with those of \citet{Geissler+2012}.

\subsubsection{V1171~Tau, HD~282954 and BD+22~574}
We observed HD~282954 and V1171~Tau two times for measurement of their proper motions with HiCIAO.
We confirmed these 3 CCs were the background stars from comparison of the astrometry between the two epochs.
Two CCs for BD+22~574 show changes in their relative distances to the central star between the two epochs, and are likely to be background stars. 
We consider it likely that the distortion correction is not perfect for the CIAO data because the distortion map for the CIAO data cannot be preperly generated due to a limited number of the field stars in Trapezium,
which was observed for the distortion correction.
However, because the distortion is small at narrow separation, the CC1 of BD+22~574 (separation $\sim$ \timeform{3''.3})  is confirmed as the background star.
It is not clear whether BD+22~574 CC2 os the companion or the background star.

\subsubsection{V855~Tau, and HD~23912}
One CC was detected for V855~Tau in January 2011.
Interestingly, however, it was not detected in January 2012.
It is difficult to conclude that we had a false detection in 2011 because it is not one of the known artifacts,
it is seen in all of the several combined images, and its PSF has a reasonable FWHM without any peculiarity in its shape.
It may therefore be a foreground object.
HD~23912 also has one CC, but it turned out to be a background star on the basis of ADI and DI observations with HiCIAO. 

\subsection{Statistical analysis for estimating the frequency of planets}
The purpose of this subsection is to constrain the frequency of planets around a star based on our observations.
First, we define and calculate the detection efficiency $\varepsilon _\mathrm{n}$ as the probability of planet detection when host star $n$ has one gas-giant planet. 

To begin with, we consider the separation range where we can detect a planet in our observations with HiCIAO/AO188.
The detection limit of a point source far from the central star is determined solely by the total integration time without being affected by the stellar halo.
As already mentioned in section \ref{sub:obs result}, the detection limit of our observations (5$\sigma$) was 20.3 magnitudes with an integration time of 5--45 minutes beyond \timeform{1''.5}. 
However, residuals of the stellar halo remain in the inner ($<$\timeform{1''.5}) region as seen in Figure~\ref{fig:detection limit of HD23912}. 
In this area, only brighter planets, brown dwarfs, and stars can be detected, but we are interested in the region where planets can be detected if they exist. 
The minimum separation for planet detection, which we define as the inner working angle (IWA), can depend not only on the sensitivity but also on the field rotation of ADI. 
In this way, the IWA is determined only by the sensitivity of our observations, which is \timeform{0''.6}--\timeform{1''.0} arcsec, depending on the amount of suppression of the stellar halo for each target. 
Nevertheless, most of the region we consider below is the outer part ($>$\timeform{1''.5}), which is free from the effect of the stellar halo. 
In the following calculation, $F_\mathrm{min}$ is defined as the minimum angular separation that a planet with a given mass $M_\mathrm{P}$ can be detected.

The $H$ magnitude can be converted to planetary mass by using the evolutionary model by \citet{Baraffe+2003}, 
assuming an age of 125~Myr and a distance of 135~pc for the Pleiades. 
Using this relation, the minimum detectable planet mass $M_\mathrm{min}$ can be determined for each separation.

Next, we calculate the detection efficiency, which is the probability that planets lie in the detectable parameter space of the observation. 
The detection efficiency $\varepsilon(M_\mathrm{P}, a, e)$ to find a planet with a certain orbit in the Pleiades is derived from the planet mass $M_\mathrm{P}$, semi-major axis $a$, and eccentricity $e$. 
Here, we assume that a host star always has one planet that has the orbital elements; 
$a$, $e$, inclination $i$ (angle between line of sight and normal to the orbital plane), and azimuth $\phi$ (angle between line of sight and periapsis). 
As the planet moves along its orbit, the separation angle $F$ from the central star to the planet varies with the true anomaly $\theta$ as described as follows:
\begin{equation}
\label{equ:1}
F=\frac{a\left( 1- e^{2}\right)}{D\left( 1 + e \cos \theta \right)} \sqrt{\cos ^{2}\left(\theta - \phi\right)\cos^{2}i + \sin^{2}\left(\theta - \phi\right)} ,
\end{equation}
where $D$ is the distance to the Pleiades cluster ($D =$135~pc). 
We then introduce $T_\mathrm{d}$, which is the time per orbital period $T_\mathrm{P}$ for a planet of $M_\mathrm{P}$ being in the range of $F\geqq F_\mathrm{min}$. 
Using $T_\mathrm{d}$, the detection efficiency of a certain orbit is described as $g(M_\mathrm{P},a,e,i,\phi)=T_\mathrm{d}/T_\mathrm{P}$. 
Considering that the line of sight is randomly distributed and independent of a planet orbit the detection efficiency for one orbit is
\begin{equation}
\label{equ:2}
\varepsilon(M_\mathrm{P},a,e)=\frac{\int^{\pi /2}_{i=-\pi /2} \sin i \int^{2\pi}_{\phi=0} g(M_\mathrm{P},a,e,i,\phi) d\phi di}{\int^{\pi /2}_{i=-\pi /2} \sin i \int^{2\pi}_{\phi=0} d\phi di}.
\end{equation}

Accordingly, the detection efficiency $\varepsilon_\mathrm{n}$ for a host star $n$ can be obtained from the distribution of planet mass, semi-major axis, and eccentricity by
\begin{equation}
\label{equ:3}
\varepsilon_\mathrm{n} = \frac{\int^{M_\mathrm{max}}_{M_\mathrm{min}}\frac{dN}{dM_\mathrm{P}} \int^{a_\mathrm{max}}_{a_\mathrm{min}} \frac{dN}{da} \int^{1}_{0}\frac{dN}{de} \varepsilon(M_\mathrm{P},a,e)dM_\mathrm{P}dade}
			{\int^{M_\mathrm{max}}_{M_\mathrm{min}}\frac{dN}{dM_\mathrm{P}} \int^{a_\mathrm{max}}_{a_\mathrm{min}} \frac{dN}{da} \int^{1}_{0}\frac{dN}{de} dM_\mathrm{P}dade} .
\end{equation}
Here, we need to consider the number distribution of planet mass, semi-major axis, and eccentricity, which are expressed as $dN/dM_\mathrm{P}$, $dN/da$ and $dN/de$, respectively. 
The distribution of planet mass was derived as $dN/dM_\mathrm{P} \propto M_\mathrm{P}^{-1.2\sim -1.9}$ by the RV survey for planets with orbital periods longer than 100 days (\cite{Cumming+2008}).
For the distribution of the semi-major axis, $dN/da \propto a^{-0.61}$ was obtained from the RV survey for planets with long orbital periods (shorter than 2000 days: \cite{Cumming+2008}). 
Finally, the distribution of eccentricity was derived as $dN/de \propto \exp(-4.2e)$ on the basis of data in The Extrasolar Planet Encyclopedia~\footnote{http://exoplanet.eu/}.
We assume these distributions in our calculation.

Adopting Baraffe et al. (2003), the minimum detectable mass in our observation was 6--10 $M_\mathrm{{J}}$ at separations larger than \timeform{1''.5}. 
As shown in Figure~\ref{fig:detection limit of HD23912}, IWAs are 100~AU for a circular orbit and 50~AU for an eccentric orbit with an eccentricity of 0.9, respectively.
Considering this result and using equation (\ref{equ:3}), the detection efficiency $\varepsilon_\mathrm{n}$ ranges from 82--96\% for a planet mass of 6--12 $M_\mathrm{{J}}$ and semi-major axis of 50--1000~AU.

In the above discussion, we calculated the detection efficiency for one planet orbiting one star, $\varepsilon_\mathrm{n}$. 
In the next step, we consider the probability of detecting {\it at least} one planet, $p_\mathrm{n}$, around a star $n$ ($n=1...N$). 
$p_\mathrm{n}$ is calculated from the detection efficiency $\varepsilon_\mathrm{n}$ and the number frequency of planets around a host star $\eta$, since
\begin{equation}
\label{equ:4}
p_\mathrm{n}=\eta \times \varepsilon_\mathrm{n} .
\end{equation}
As noted above, $\varepsilon_\mathrm{n}$ is uniquely determined by the orbital distribution of a planet and the detection separation range in the observations. 
On the other hand, $p_\mathrm{n}$ can be constrained by our imaging results for 20 stars.
Therefore, it is possible to constrain the planet frequency $\eta$ for a host star.

In the following analytical approach, we employ Bayes' theorem as described by \citet{Vigan+2012} and \citet{Lafreniere+2007}. 
The probability of detecting at least one planet is $\eta \times \varepsilon_\mathrm{n}$ while that of non-detection is $(1-\eta \times \varepsilon_\mathrm{n})$. 
The likelihood of the data given $\varepsilon_\mathrm{n}$ is described as
\begin{equation}
\label{equ:5}
L(\{d_\mathrm{n}\}|\eta)= \prod^{N}_{n=1}\left(1-\eta \varepsilon_\mathrm{n}\right)^{1-d_\mathrm{n}}\cdot \left( \eta \varepsilon_\mathrm{n} \right)^{d_\mathrm{n}} ,
\end{equation}
where $d_\mathrm{n}$ is the sign of detection, which equals 1 if at least one planet is detected around a star n and 0 if no planet is detected. 
On the left-hand side of the equation, $\{d_\mathrm{n}\}$ shows the set of results from $N$ observations. 
Using this likelihood, the conditional probability distribution that the set of events \{$d_\mathrm{n}$\} occurs with frequency $\eta$ is 
\begin{equation}
\label{equ:6}
p\left( \eta | \{d_\mathrm{n}\} \right) = \frac{L(\{d_\mathrm{n}\}|\eta)p(\eta)}{\int^1_{0} L(\{d_\mathrm{n}\}|\eta)p(\eta) d\eta} ,
\end{equation}
where p($\eta$) is the prior probability of $\eta$. 
Since $\eta$ is unknown a priori, $p(\eta)=1$.

We can determine the range of $\eta$ as a confidence interval (CI) on a given confidence level (CL) $\alpha$,
\begin{equation}
\label{equ:7}
\alpha = \int^{\eta_\mathrm{max}}_{\eta_\mathrm{min}} p\left( \eta | \{d_\mathrm{n}\} \right) d\eta ,
\end{equation}
where $\eta_{max}$ and $\eta_{min}$ are the maximum and minimum values of $\eta$ in the case of \{$d_\mathrm{n}$\}.

In our observations, there are 8 companion candidates without proper motion measurements (V1171~Tau CC1, CC2, BD+22~574 CC2, HD~282954 CC1, V1054~Tau CC1, CC2,  V1174~Tau CC1, and CC2). 
Even if they are companion objects, their masses are larger than that of planets ($> 12 M_\mathrm{{J}}$).
Thus, no planets is found around 20 stars in our observations, resulting in a value of $\eta_\mathrm{max}$ of about 17.9\% (CL = 95\%) for planets in the mass range of 6--12 $M_\mathrm{{J}}$ and the semi-major axis of 50--1000~AU. 
The minimum value $\eta_\mathrm{min}$ is always 0 in this case.

\section{Discussion}
On the basis of our observations of 20 stars, the frequency of planets in the mass range of 6--12 $M_\mathrm{{J}}$ orbiting at a distance of 50--1000~AU from a host star in the Pleiades (125~Myr, 135~pc) is estimated to be 17.9\% as an upper limit ($2\sigma$). 
This is the first time this constraint has been obtained for a certain age ($\sim$125~Myr). 

In a previous direct imaging survey by \citet{Lafreniere+2007}, the frequency of planets over the mass and separation ranges of 0.5--13 $M_\mathrm{{J}}$ and 50--250~AU was below 10\%, as derived from observations of 85 stars with the Gemini North telescope. 
Similarly, the frequency of planets of $>1~M_\mathrm{J}$ at 40--500~AU was not greater than 9.3\% ($2\sigma$) by the VLT observations of 88 stars within 100~pc (\cite{Chauvin+2010}).
Therefore, our estimate is consistent with these previous results, indicating that the planet frequency in the Pleiades is not much higher than in other moving groups and around field stars. 

According to these results, giant planets are very rare at larger separations (more than about 50~AU), although there are a few known candidate systems (e.g., \cite{Marois+2008, Itoh+2005}). 
Since current formation theory predicts that heavy giant planets can form only via disk instability at distant regions, 
it is speculated that such instability is not a major in-situ formation process for giant planets.
Furthermore, our observations cover a wide area even beyond a few 100~AU which is the typical size of protoplanetary disks (\cite{AndrewWilliams2007}), 
thus it is difficult to expect that planets form in situ at such a distances from a host star. 
However, it has been suggested that giant planets or their natal fragments in multiple planetary systems can be ejected into very wide orbits ($10^{2}$--$10^5$~AU) through gravitational interaction (\cite{BasuVorobyov2012, Veras+2009}). 
At present, the observed rareness is not inconsistent with theoretical predictions that invoke planet--planet scattering. 

\begin{figure}
  \begin{center}
   \FigureFile(80mm,80mm){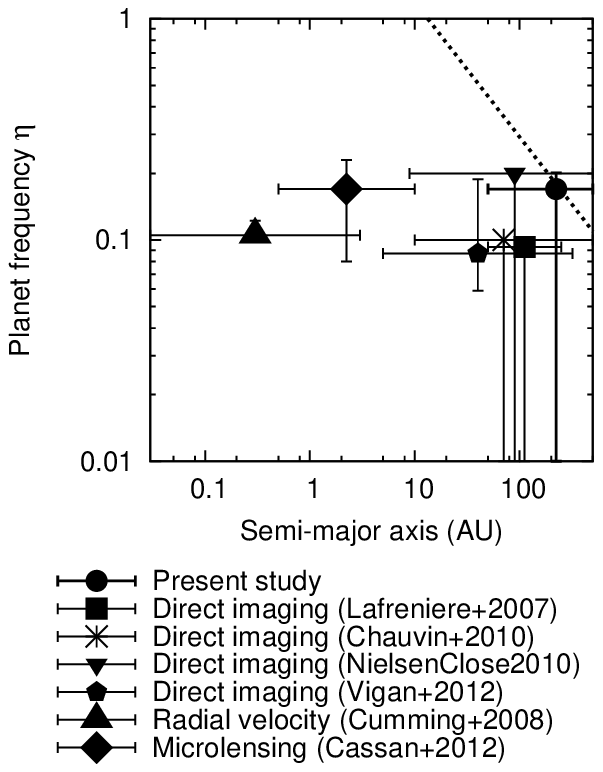}
   \FigureFile(80mm,80mm){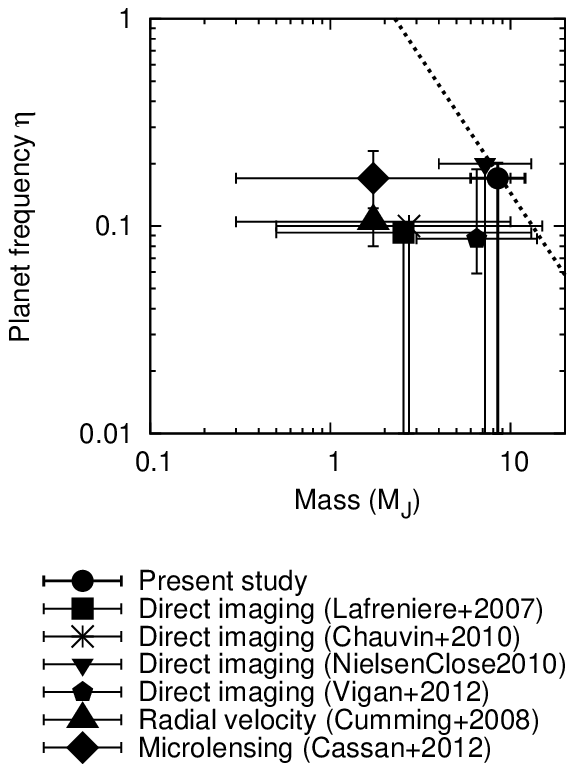}
  \end{center}
  \caption{
  Planet frequency $\eta$ as a function of semi-major axis ({\it left panel}) and planet mass ({\it right panel}).
  {\it The circle} shows our work (50--1000~AU, 6--12 $M_\mathrm{J}$), while {\it the square} indicates the direct imaging (50--250~AU, 0.5--13 $M_\mathrm{J}$; \cite{Lafreniere+2007}). 
  {\it The rice symbol}, {\it the triangle} and {\it the pentagon} shows other direct imaging (10--500~AU, 0.5--15 $M_\mathrm{J}$; \cite{Chauvin+2010}), 
  (8.9--911~AU, $>$4 $M_\mathrm{J}$; \cite{Nielsen+2010}) and (5--320~AU, 3--14 $M_\mathrm{J}$; \cite{Vigan+2012}), respectively.
  {\it The triangle} denotes the radial velocity (0.03--3~AU, 0.3--10 $M_\mathrm{J}$; \cite{Cumming+2008}), and {\it the diamond} shows microlensing (0.5--10AU, 0.3--10 $M_\mathrm{J}$; \cite{Cassan+2012}). 
  The dotted lines in the two panels indicate the distribution of the frequency of planets that is derived from our observations. 
  The slopes of the lines are $-1.31$ and $-0.61$ in the $left$ and $right$ panels, respectively.
   }\label{fig:smadistri}
\end{figure}

In other planet surveys of the region near host stars using microlensing (OGLE: \cite{Beaulieu+2006, Kubas+2008}, MOA: \cite{Sumi+2010}), the frequency of planets with 0.3--10~$M_\mathrm{{J}}$ at 0.5--10~AU was $17^{+6}_{-9}$\% (\cite{Cassan+2012}). 
In addition, the frequency of planets more massive than 0.3--10~$M_\mathrm{{J}}$ over 0.03--3~AU was 10.5 $\pm 1.7$\% by RV survey (\cite{Cumming+2008}). 
Though the detectable separation in these other surveys was different from that in direct imaging, the frequency of planets according to our survey does not seem to be higher than those obtained by microlensing and RV surveys (Figure~\ref{fig:smadistri}, Table~\ref{tab:summary of obs.}).

\begin{table*}
  \caption{Comparison of observations for planet frequency.}\label{tab:summary of obs.}
  \begin{center}
    \begin{tabular}{lcccc}
      \hline
      Observation		&	Ref.					& \multicolumn{2}{c}{Distribution index}								&	Planet frequency		\\
      method			&						& Mass ($\alpha$)					& Semi-major axis ($\beta$)	&	 ($\eta$)				\\
      					&						& $dN/dM_\mathrm{P}\propto (M_\mathrm{P})^{\alpha}$	& $dN/da\propto a^{\beta}$	&	 					\\
      \hline
      Direct Imaging	& Present work			& -1.31 								& -0.61						& $\le 17.9$ \%			\\
      Direct Imaging	& \cite{Lafreniere+2007} 	& -1.2								& -1.0						& $\le 9.3$ \%			\\
      Radial velocity		& \cite{Cumming+2008}	& -1.31								& -0.61 						& $10.5 \pm 1.7$ \%		\\
      Microlensing		& \cite{Cassan+2012}		& -1.68								& -1.0 						& $17^{+6}_{-9}$ \%		\\
      \hline
      \multicolumn{5}{@{}l@{}}{\hbox to 0pt{\parbox{160mm}{\footnotesize
      \par\noindent
      		Our use of $\beta$ is taken from \citet{Cumming+2008}. 
		In direct imaging by \citet{Lafreniere+2007}, $\alpha$ and $\beta$ were the values extrapolated from RV observations.
		\citet{Lafreniere+2007} and \citet{Cassan+2012} assumed a flat distribution in logarithmic semi-major axis space.
		\par\noindent
     }\hss}}
    \end{tabular}
  \end{center}
\end{table*}

In our observations, point sources fainter than 14.5 mag are detected around 9 of 20 (40\%) target stars whether or not they are real companion objects. 
The detection limit is 20.3 mag in the $H$ band at the separation of \timeform{1''.5}--\timeform{10''}. 
This possibility of finding other point sources is consistent with previous direct imaging studies with similar survey depth and size of the field of view. 
For instance, CCs were detected toward 32 stars (44\%) in the galactic latitude of $>|10|$ degrees in the imaging by \citet{Chauvin+2010}. 
Among them, 5 stars have already been confirmed as background objects while 78\% remain to have their proper motion observed. 
It is highly likely that most of them are background stars, but we would like to point out that as a by-product, deep direct imaging would also be useful to discuss galactic models.
This, however, is beyond the scope of our paper. 

\section{Summary}
We have carried out a SEEDS imaging survey for detection of extrasolar gas-giant planets in the Pleiades with the near-infrared imaging instrument HiCIAO and the adaptive optics instrument AO188 on the Subaru telescope between October 2009 and January 2012. 
Thirteen companion candidates were found around 9 host stars in $H$ band by using ADI observations. 
The detection limit of our observations (5$\sigma$) was 20.3 magnitudes with an integration time of 5--45 minutes beyond \timeform{1''.5}. 
For HD~23514 and HII~1348, we confirmed a brown dwarf respectively, which were detected by a previous study with proper motion measurement (\cite{Rodriguez+2012, Geissler+2012}). 
Five of the 13 candidates were confirmed to be background stars on the basis of proper motion. 
One was not found in the second epoch observation; thus, this was unlikely to be a background or companion object.
Only one it was not confirmed whether or not it is background star, as the precision of their proper motions was not sufficient.
Four of the 13 remain to be observed to confirm whether they are co-moving.

We determined the detection efficiency, which is the probability of finding a 6--12 Jovian-mass planet at 50--1000~AU from the host star in the Pleiades, to be about 90\% on the basis of our detection limit.
Because there was no detection of such a planet, we estimated that the frequency of stars having gas-giant planets in the Pleiades is less than 17.9\%. 
This result is consistent with previous direct imaging studies, indicating that planet frequency in the Pleiades is not considerably higher than those obtained in moving groups and field stars.

\bigskip

\appendix
\section*{Observation images}

Details regarding the images and reduction are described in Table~\ref{tab:summary of observations} and section~\ref{sec:data reduction}.
All images are obtained through ADI reduction in the $H$ and $K_\mathrm{S}$ (only Figure~\ref{fig:TYC1800}; TYC~1800-2144-1).
The field of view of all images is \timeform{19''.5} $\times$ \timeform{19''.5}.
The circle in images represents the position of the companion candidates (CCs).

\begin{figure*}
  \begin{center}
  \subfigure[BD+22 574]{
    \FigureFile(80mm,80mm){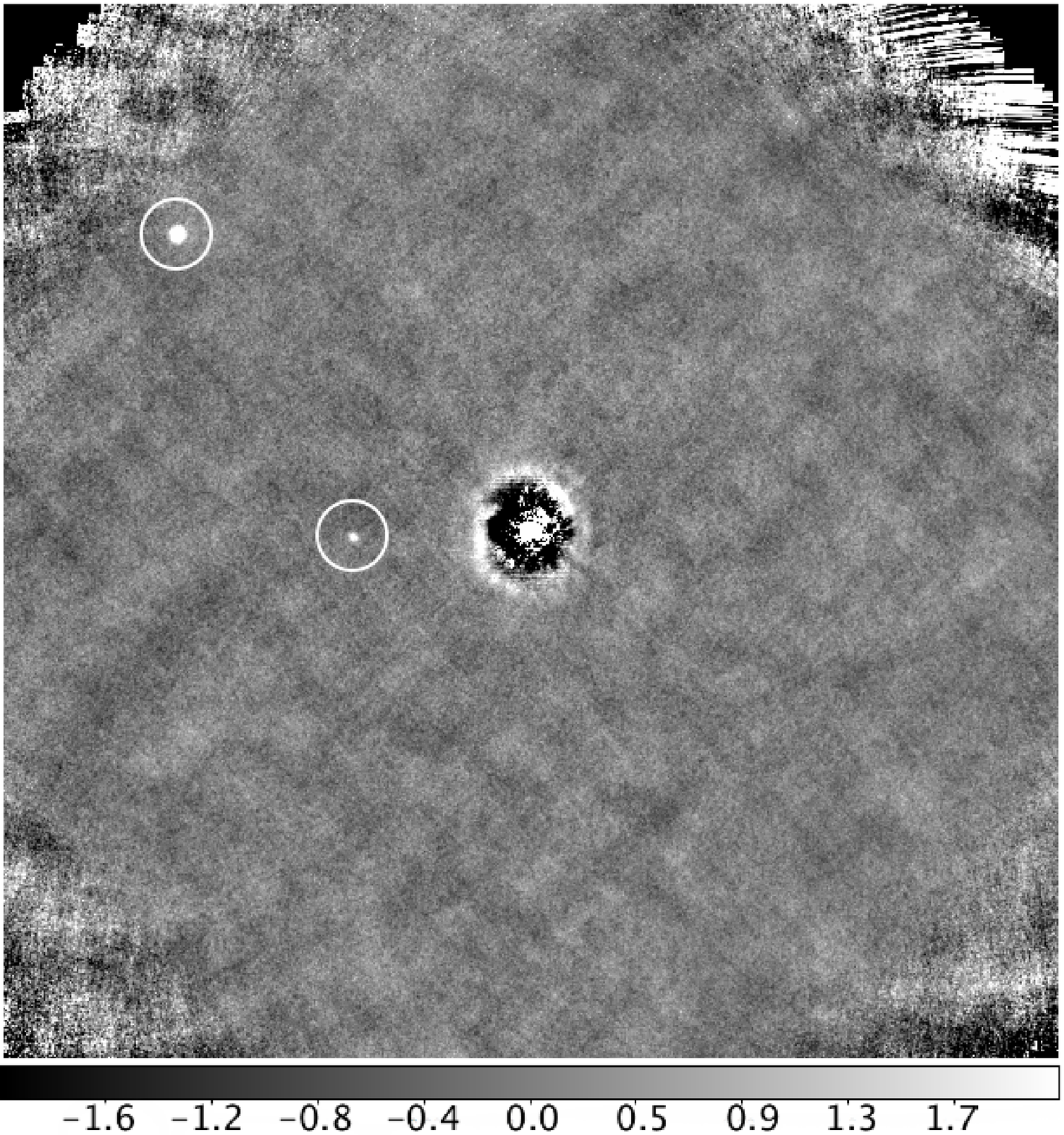}
    \label{fig:BD+22 574}}
  \subfigure[V1171~Tau]{
     \FigureFile(80mm,80mm){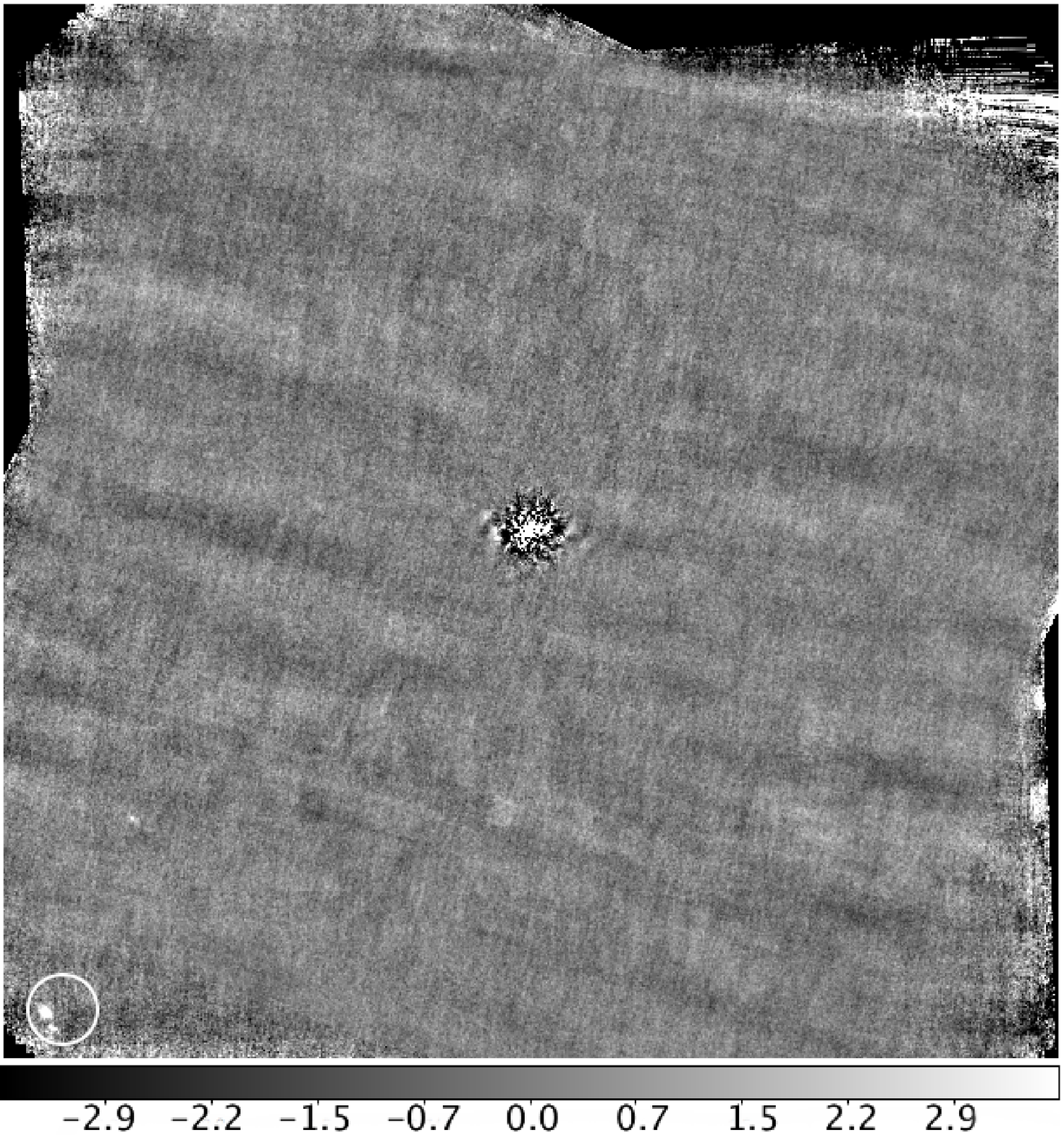}
     \label{fig:V1171Tau}}
  \end{center}

  \begin{center}
  \subfigure[HII~2462]{
    \FigureFile(80mm,80mm){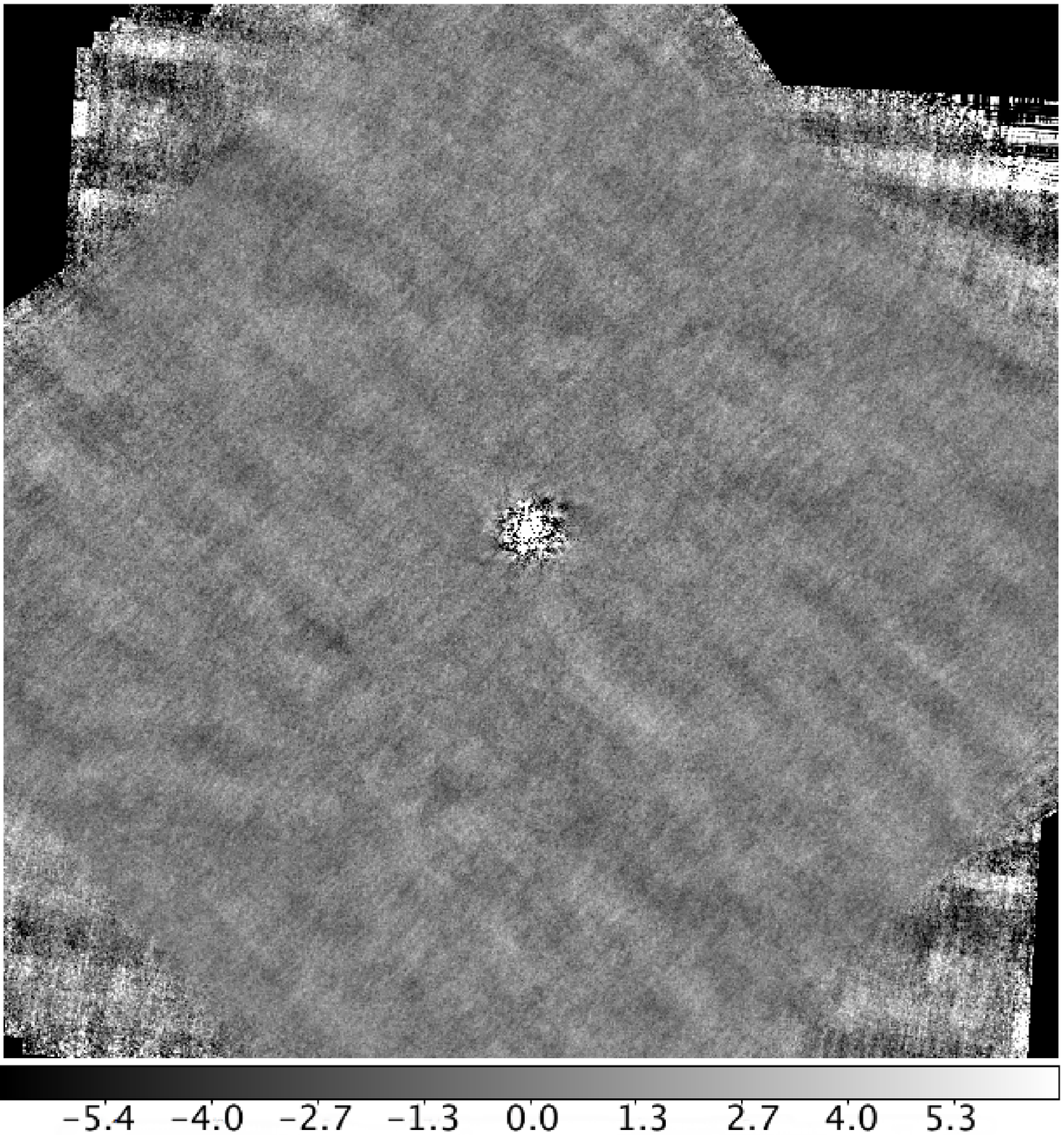}
    \label{fig:HII 2462}}
  \subfigure[HD~23863]{
    \FigureFile(80mm,80mm){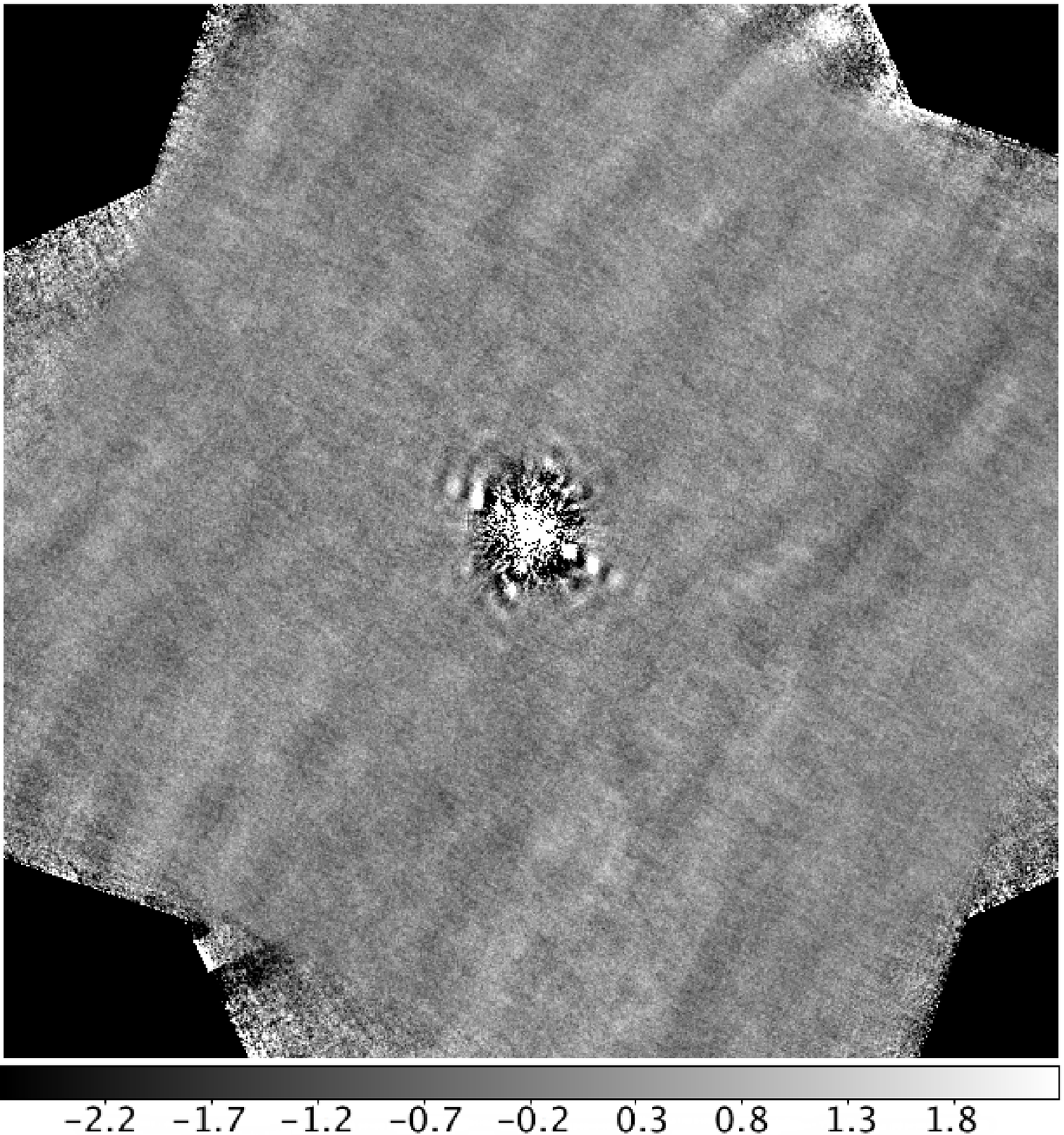}
    \label{fig:HD23863}}
  \end{center}
  \caption{
  	{\it Top left panel:} BD+22~574.  
	{\it Top right panel:} V1171~Tau. 2 CCs are in one circle.
	{\it Lower left panel:} HII~2462.
 	{\it Lower right panel:} HD~23863.
	The unit of the color bar is ADU per each exposure time.
	}\label{fig:all results1}
\end{figure*}

\addtocounter{figure}{-1}
\begin{figure*}
\addtocounter{subfigure}{4}
  \begin{center}
  \subfigure[HD~23912 (2010)]{
    \FigureFile(80mm,80mm){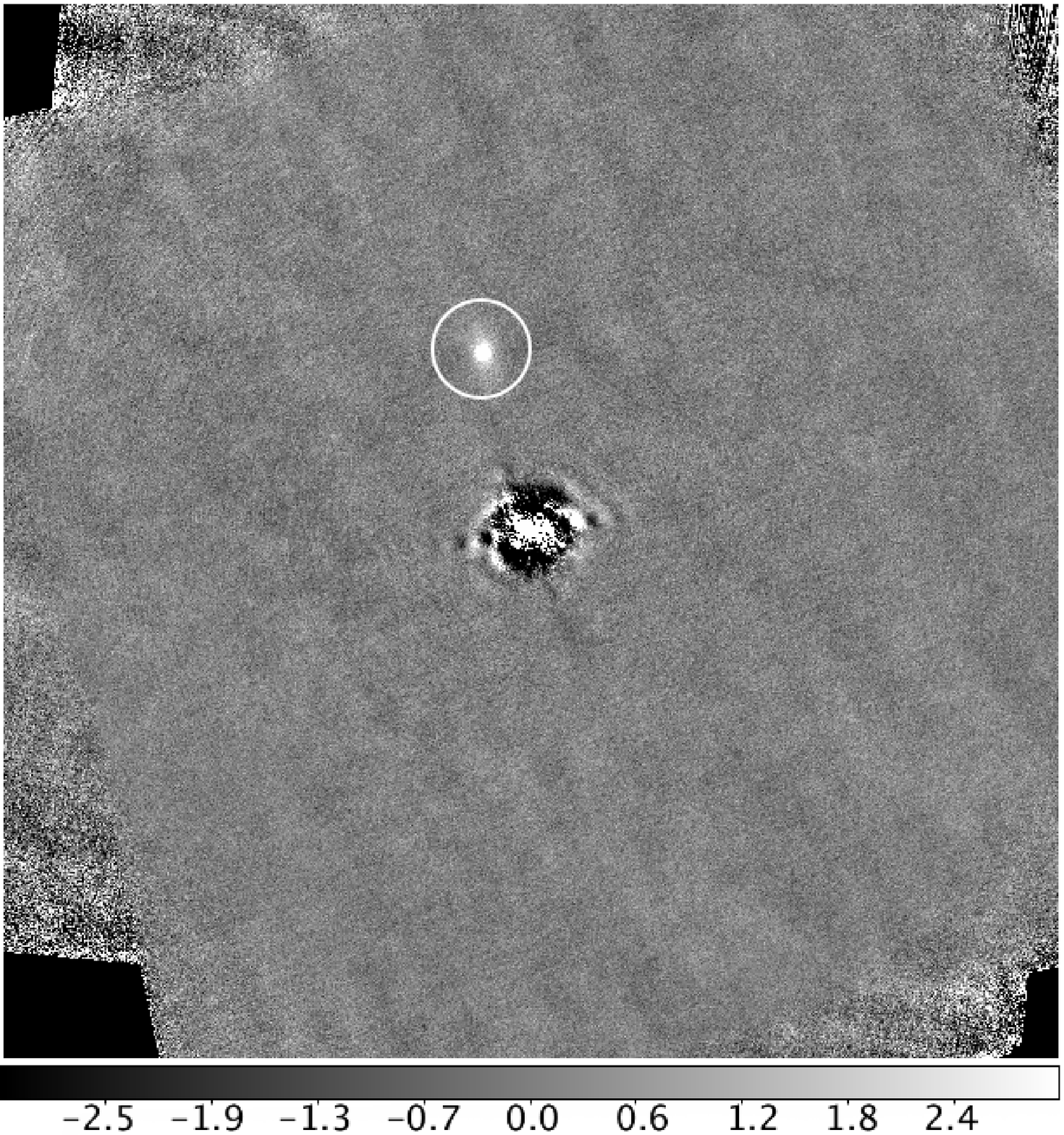}
    \label{fig:HD23912_2010}}
  \subfigure[HD~282954]{
    \FigureFile(80mm,80mm){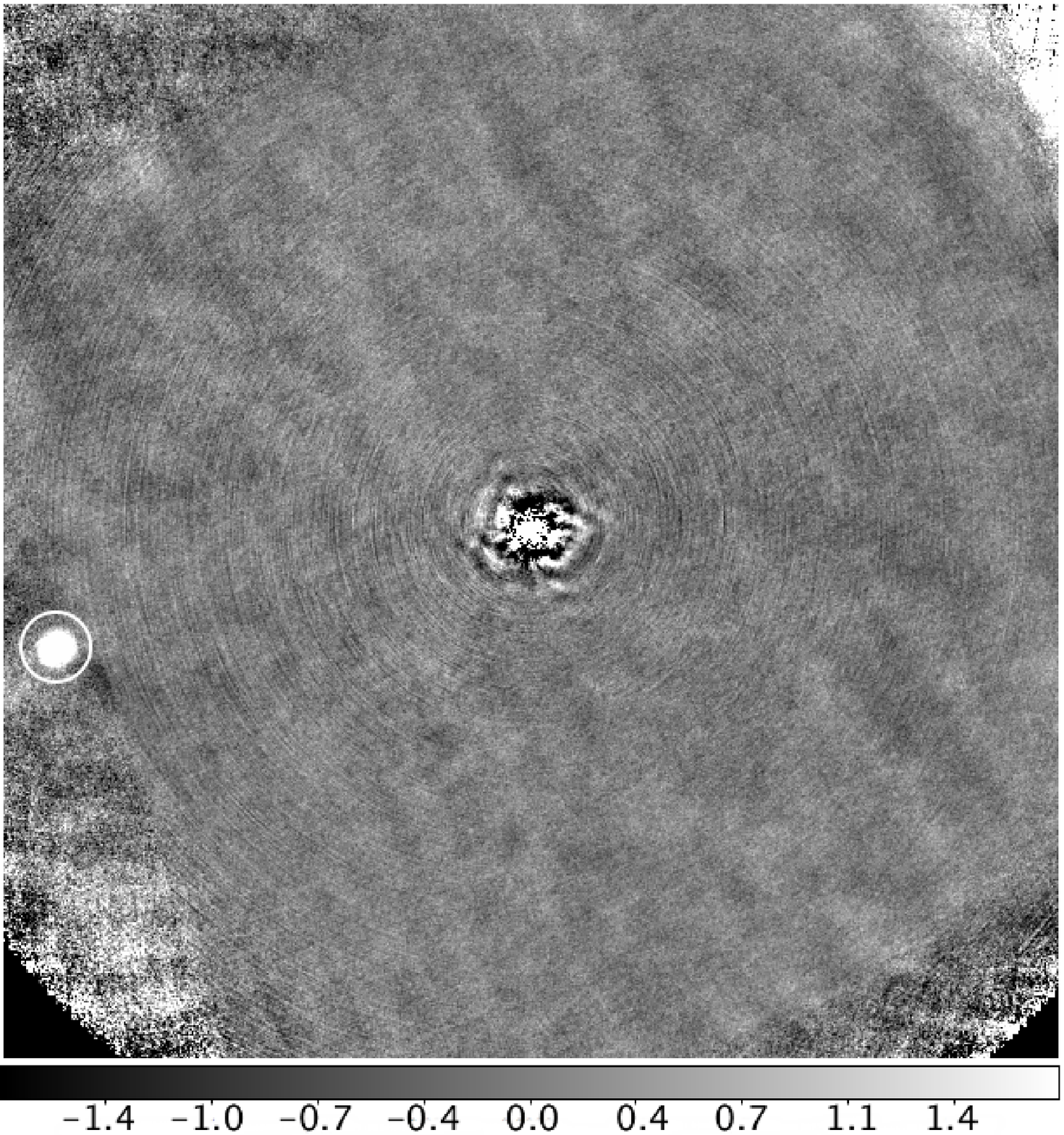}
    \label{fig:HD282954}}
  \end{center}
  \begin{center}
  \subfigure[HD~23514]{
     \FigureFile(80mm,80mm){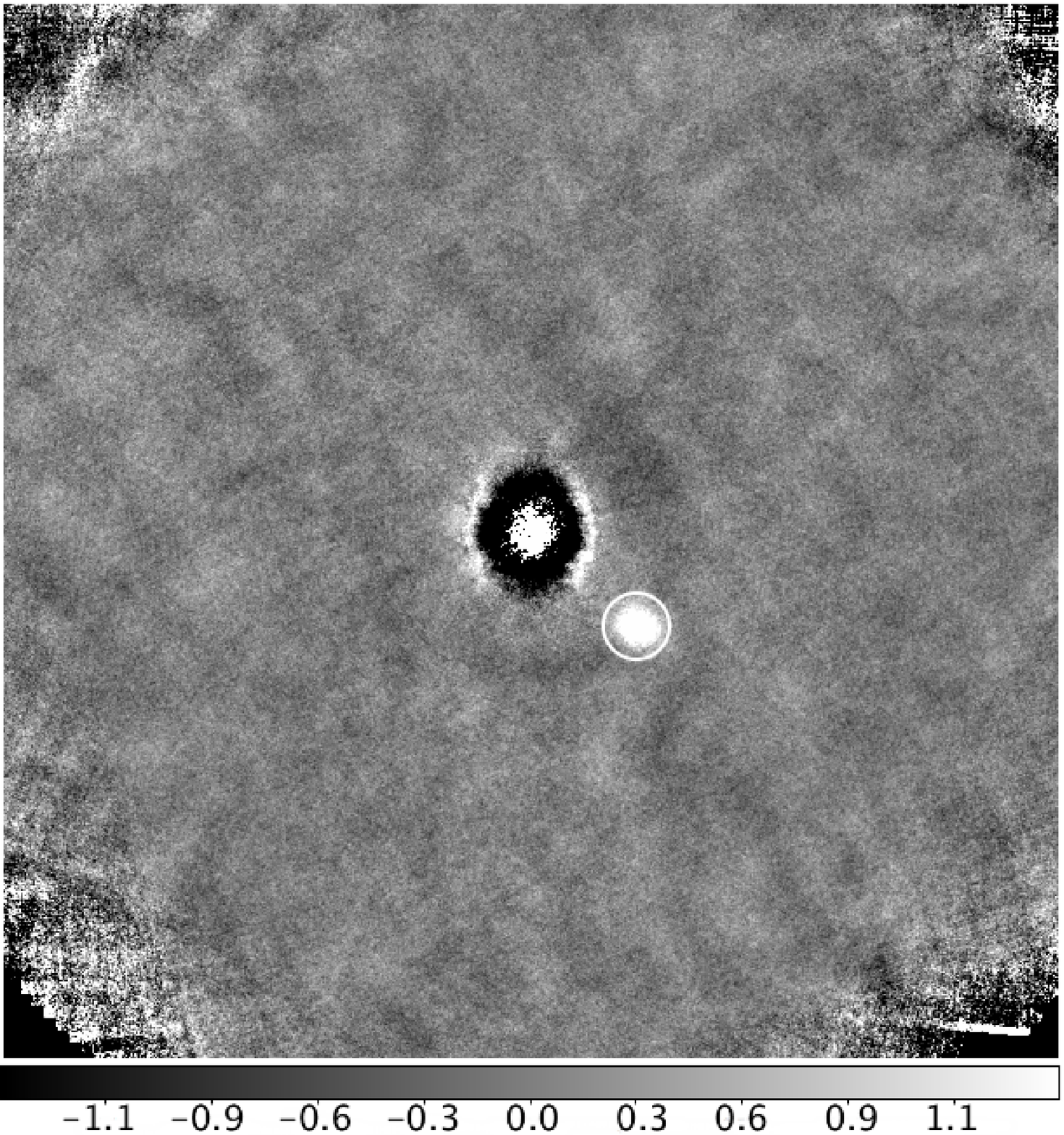}
     \label{fig:HD23514}}
  \subfigure[HD~23247 (2011)]{
    \FigureFile(80mm,80mm){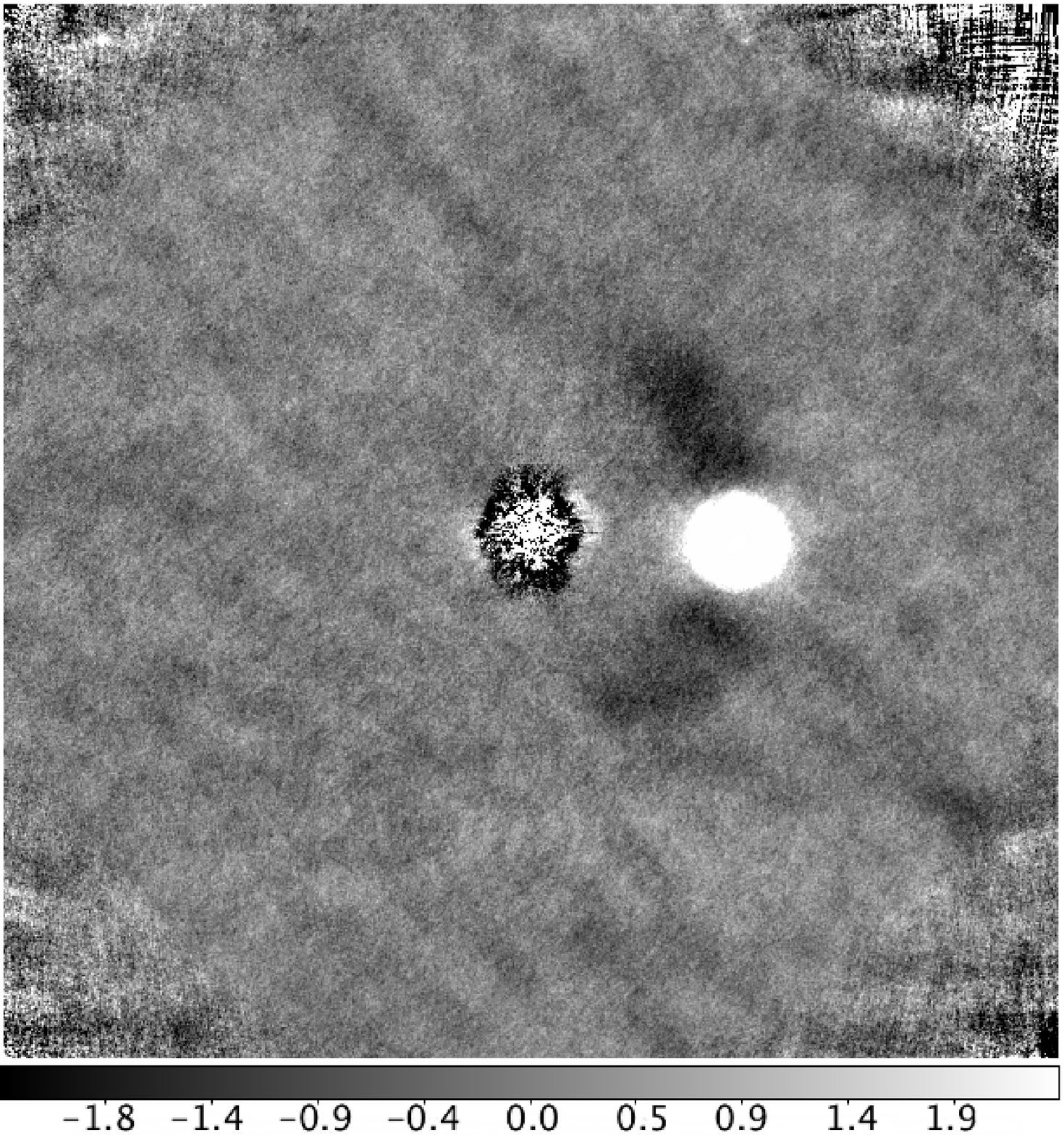}
    \label{fig:HD23247(201101)}}
  \end{center}
  \caption{
  {\it Continued.}
		\par\noindent
  		{\it Top left panel:} HD~23912~(2010).
   		{\it Top right panel:} HD~282954.	
  		{\it Lower left panel:} HD~23514.
		{\it Lower right panel:} HD~23247 (2011). 
		The unit of the color bar is ADU per each exposure time.
		}\label{fig:all results2}
\end{figure*}

\addtocounter{figure}{-1}
\begin{figure*}
\addtocounter{subfigure}{8}
  \begin{center}
  \subfigure[V855~Tau (2011)]{
    \FigureFile(80mm,80mm){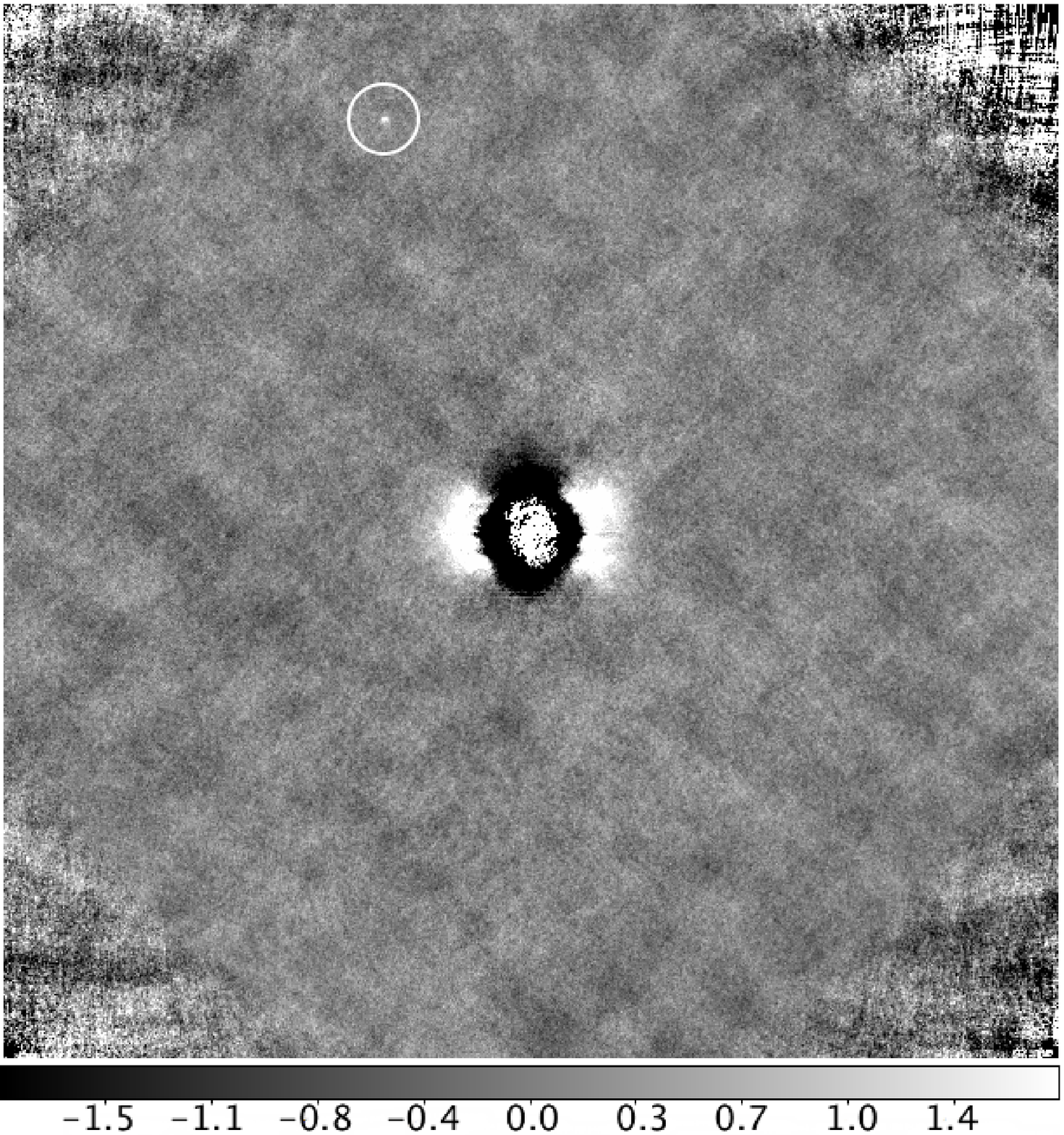}
    \label{fig:V855Tau(2011)}}
  \subfigure[HD~24132]{
    \FigureFile(80mm,80mm){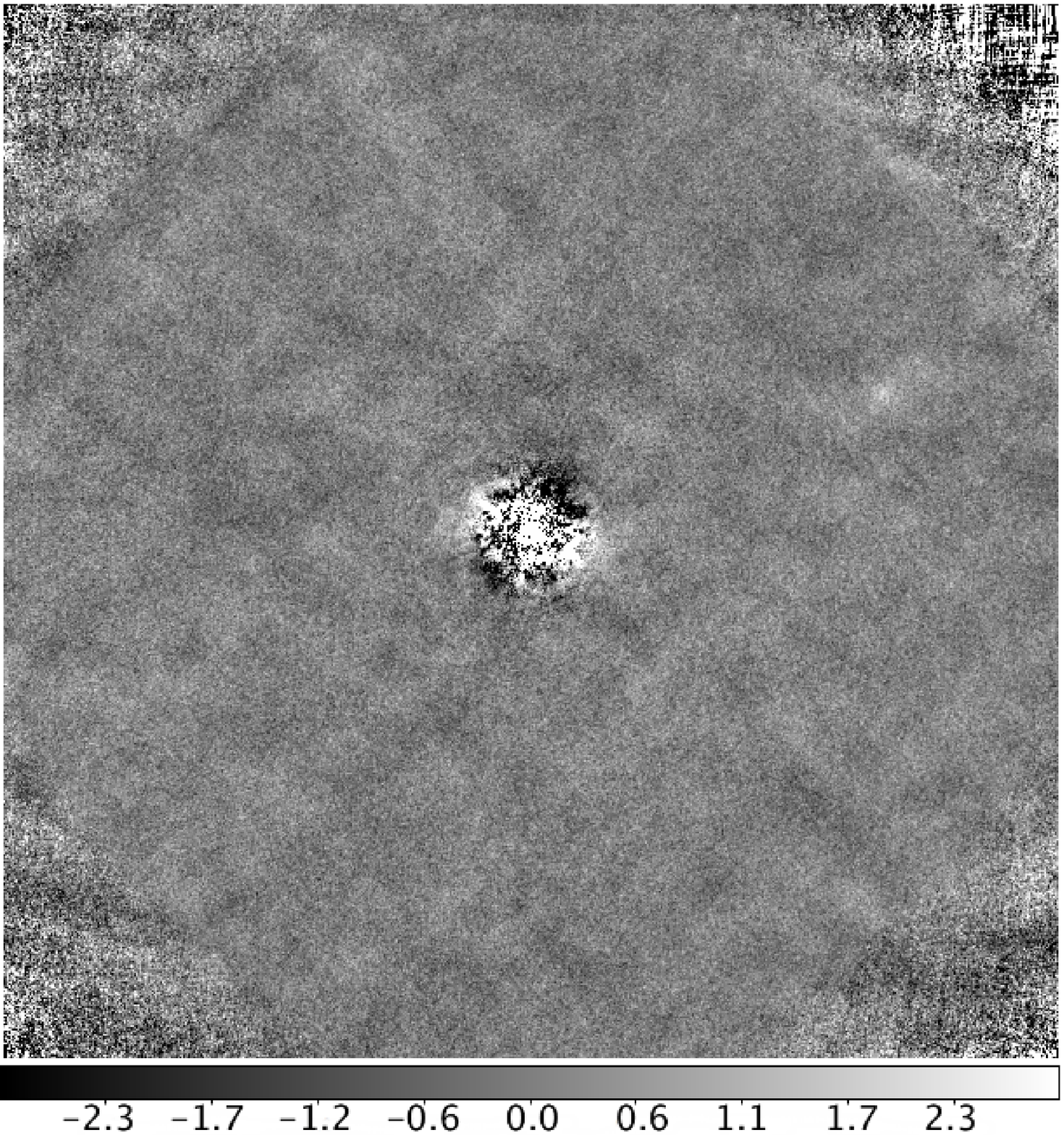}
    \label{fig:HD24132}}
 \end{center}
  \begin{center}
  \subfigure[HD~23061]{
    \FigureFile(80mm,80mm){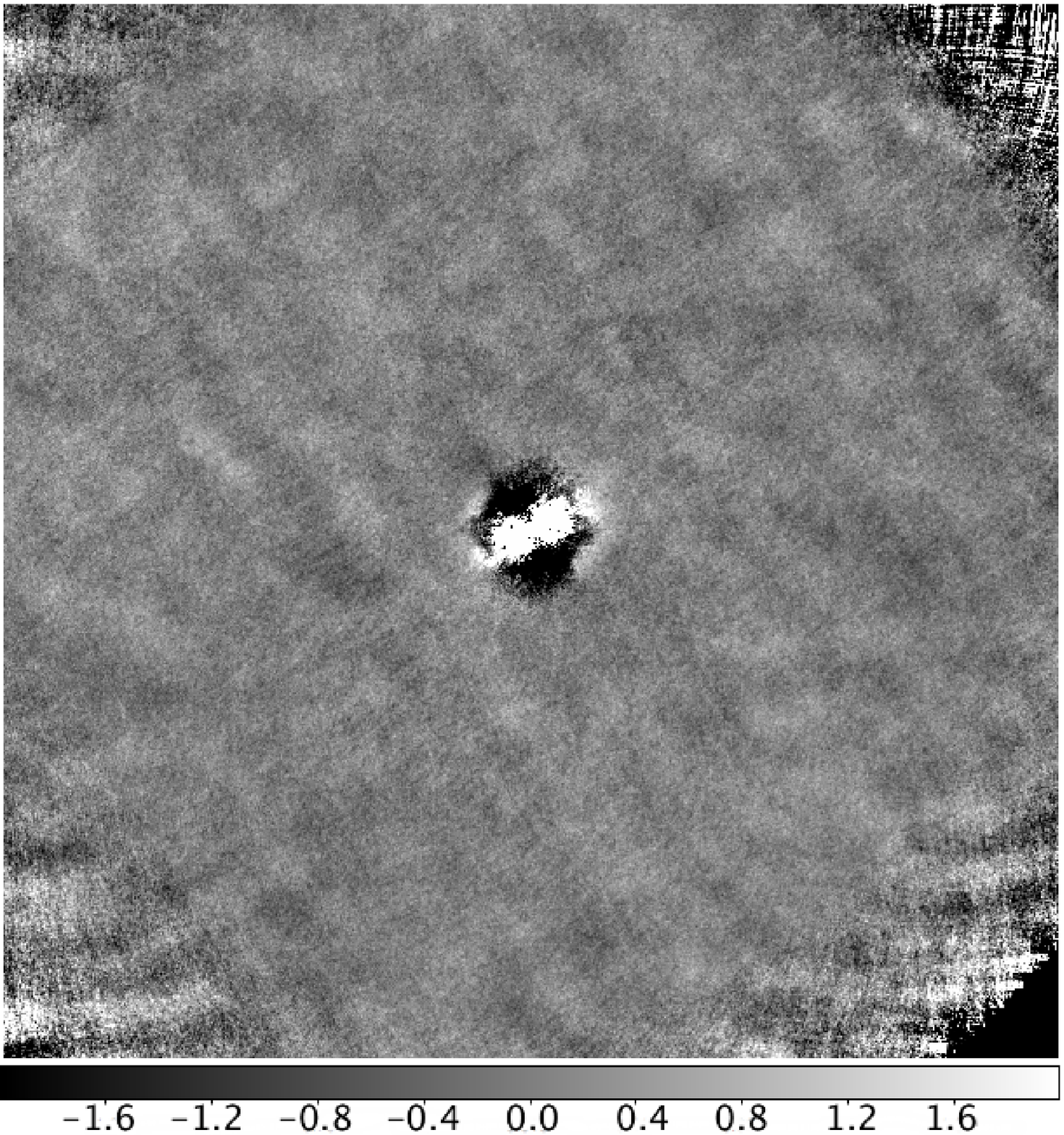}
    \label{fig:HD23061}}
  \subfigure[TYC~1800-2144-1]{
    \FigureFile(80mm,80mm){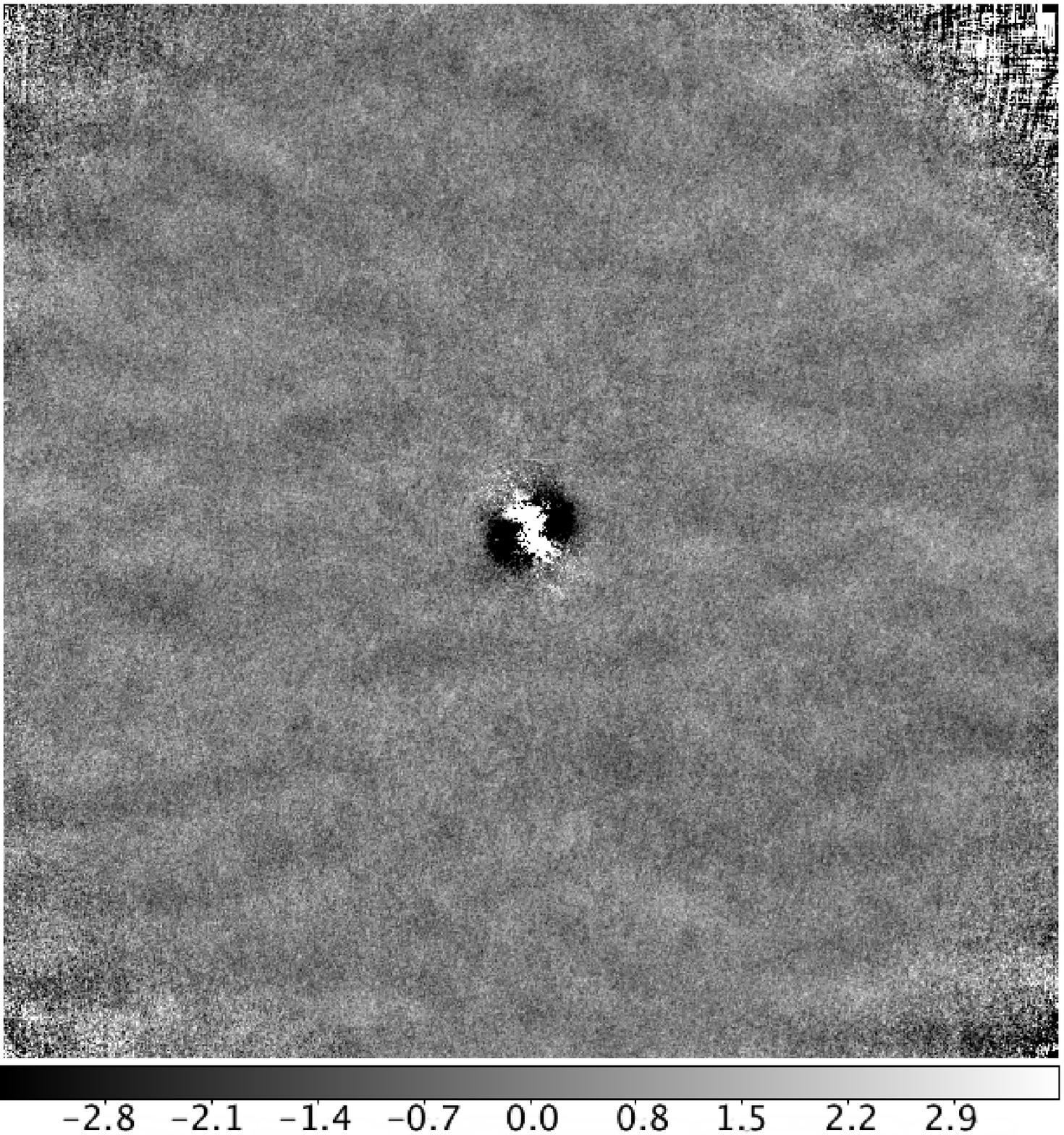}
    \label{fig:TYC1800}}
  \end{center}
  \caption{
  {\it Continued.}
		\par\noindent
  		{\it Top left panel:} V855~Tau (2011).
		{\it Top right panel:} HD~24132.
  		{\it Lower left panel:} HD~23061.
 		{\it Lower right panel:}TYC~1800-2144-1.	
		The unit of the color bar is ADU per each exposure time.
		}\label{fig:all results3}
\end{figure*}

\addtocounter{figure}{-1}
\begin{figure*}
\addtocounter{subfigure}{12}
  \begin{center}
  \subfigure[HII~1348]{
     \FigureFile(80mm,80mm){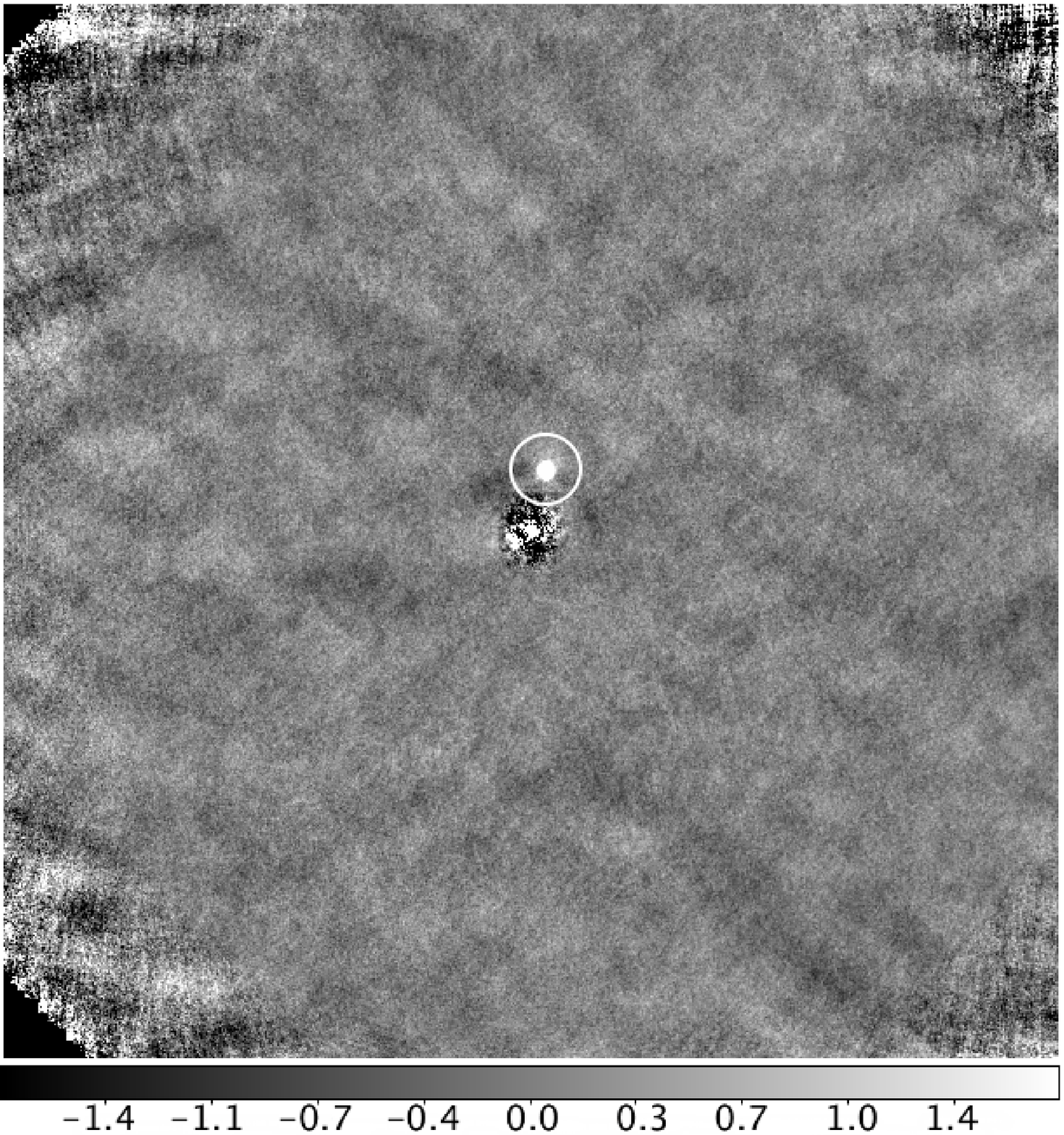}
     \label{fig:HII1348}}
  \subfigure[Melotte~22~SSHJ~G214]{
    \FigureFile(80mm,80mm){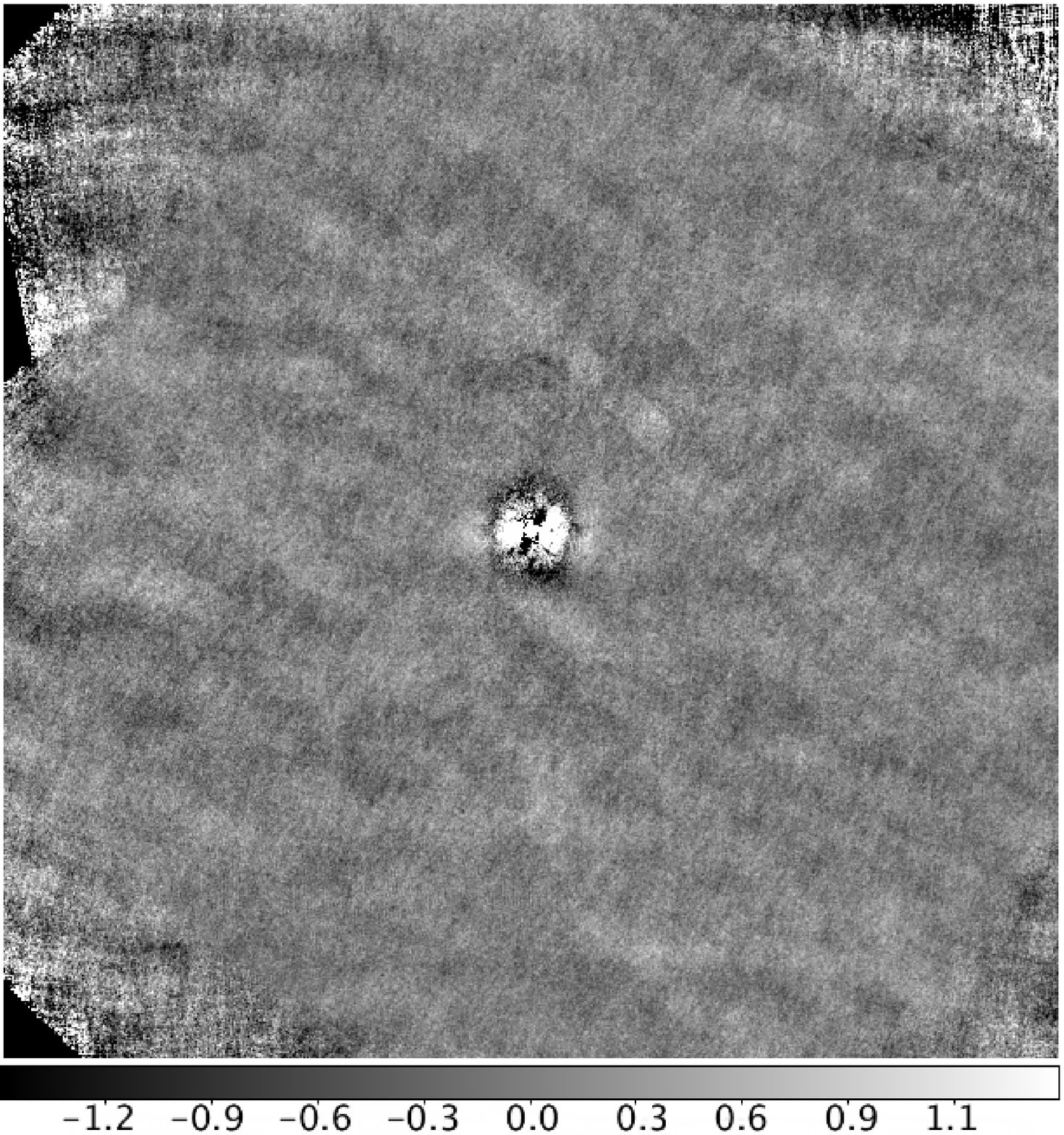}
    \label{fig:G214}}
  \end{center}
  \begin{center}
  \subfigure[BD+23~514]{
    \FigureFile(80mm,80mm){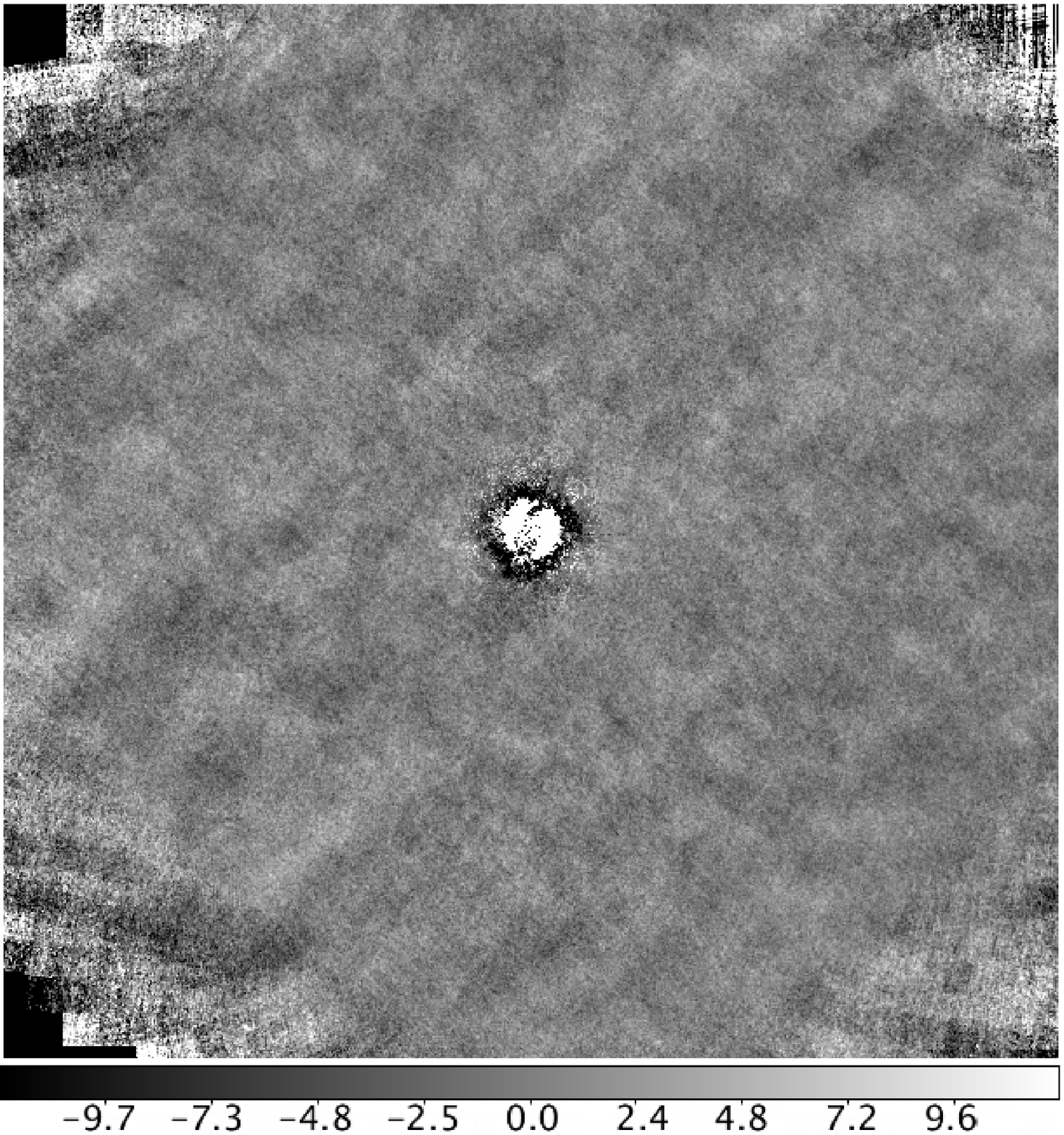}
    \label{fig:BD+23 514}}
  \subfigure[Melotte~22~SSHJ~G213]{
    \FigureFile(80mm,80mm){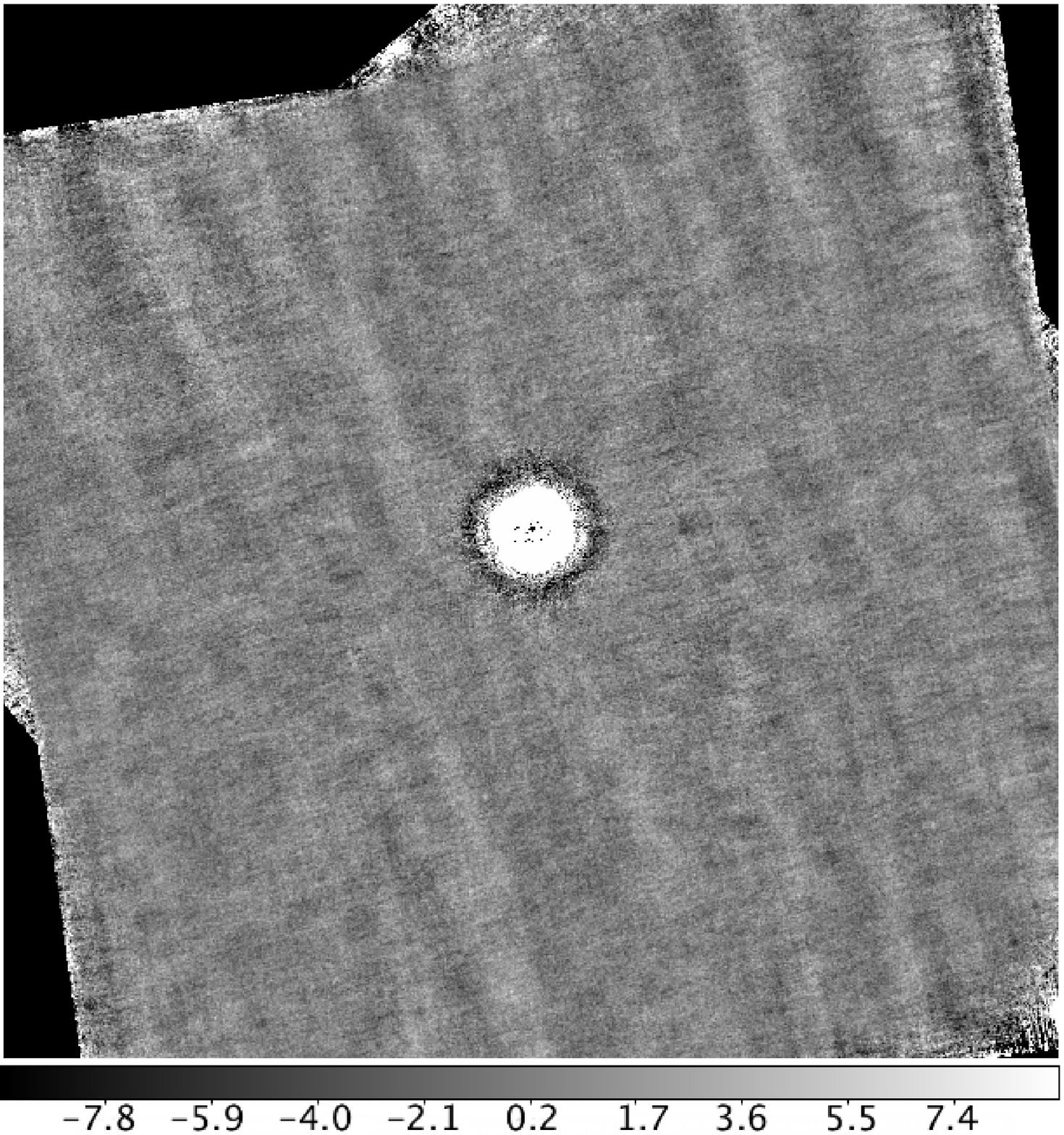}
    \label{fig:G213}}
  \end{center}
  \caption{
  {\it Continued.}
		\par\noindent
  		{\it Top left panel:} HII~1348.
		{\it Top right panel:} Melotte~22~SSHJ~G214.
  		{\it Lower left panel:} BD+23~514.
  		{\it Lower right panel:} Melotte~22~SSHJ~G213.
		The unit of the color bar is ADU per each exposure time.
		}\label{fig:all results4}
\end{figure*}

\addtocounter{figure}{-1}
\begin{figure*}
\addtocounter{subfigure}{16}
  \begin{center}
  \subfigure[Melotte~22~SSHJ~G221]{
     \FigureFile(80mm,80mm){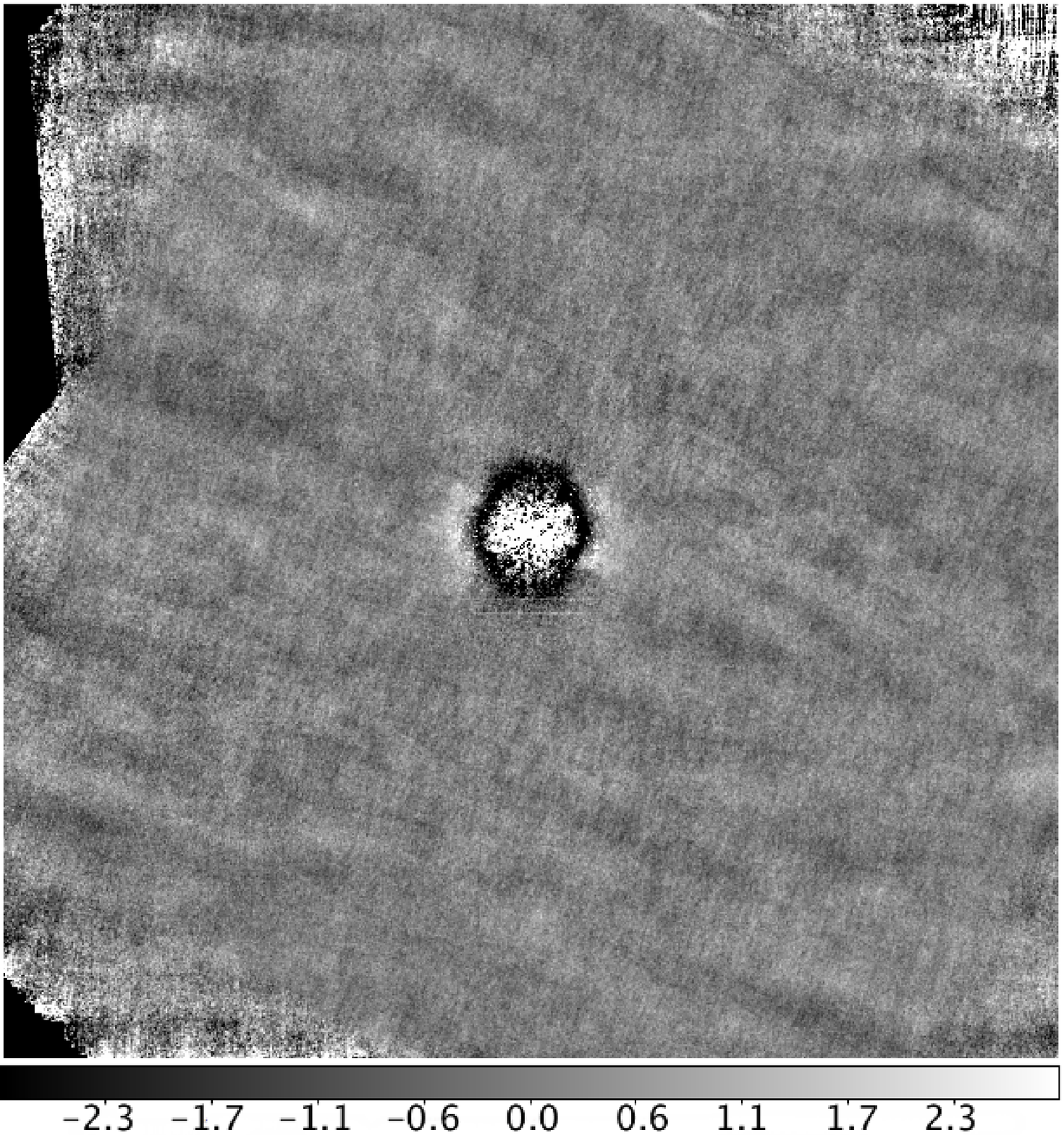}
     \label{fig:G221}}
  \subfigure[V1054~Tau]{
    \FigureFile(80mm,80mm){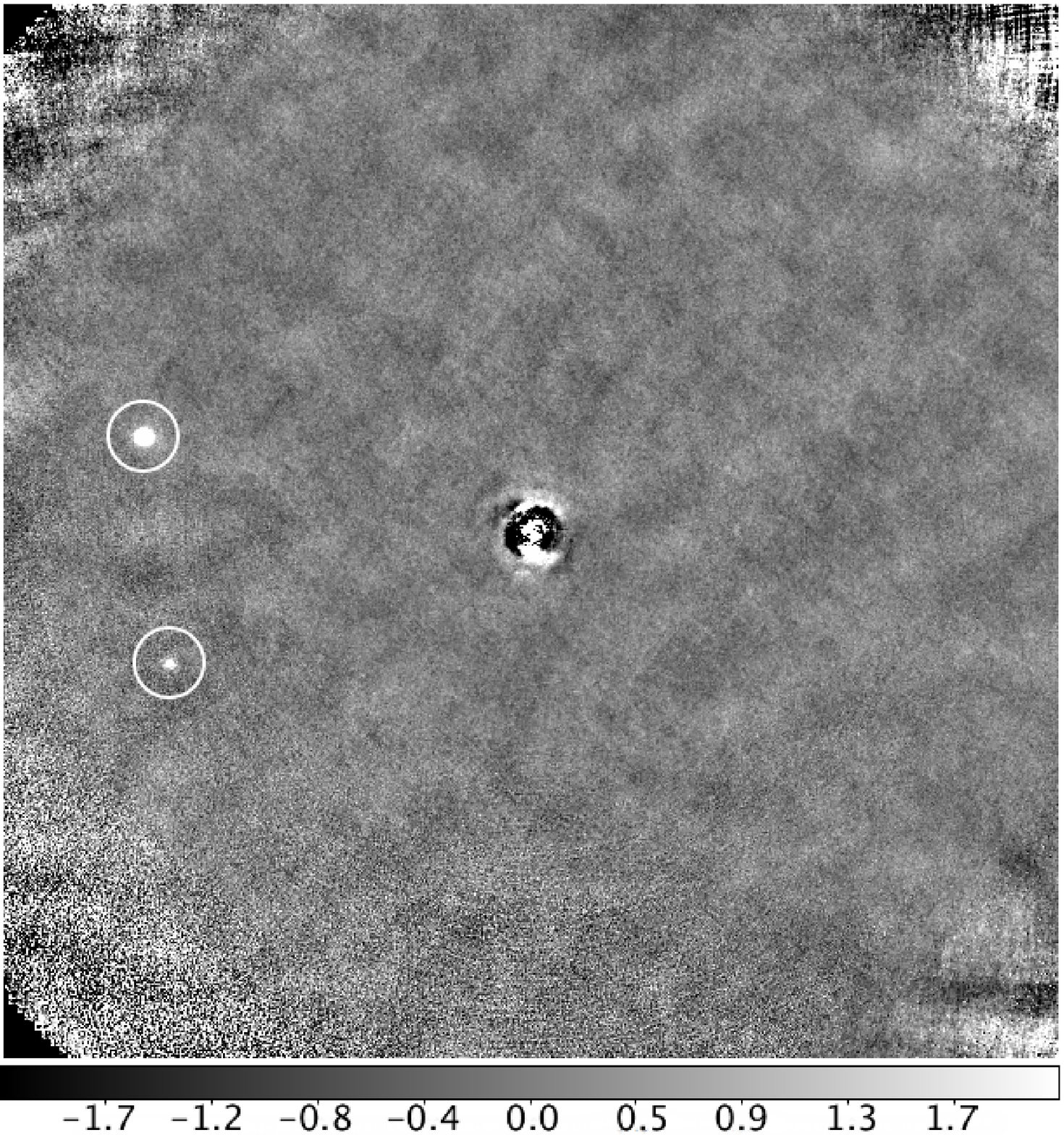}
    \label{fig:V1054Tau}}
  \end{center}
  \begin{center}
  \subfigure[V1174~Tau]{
    \FigureFile(80mm,80mm){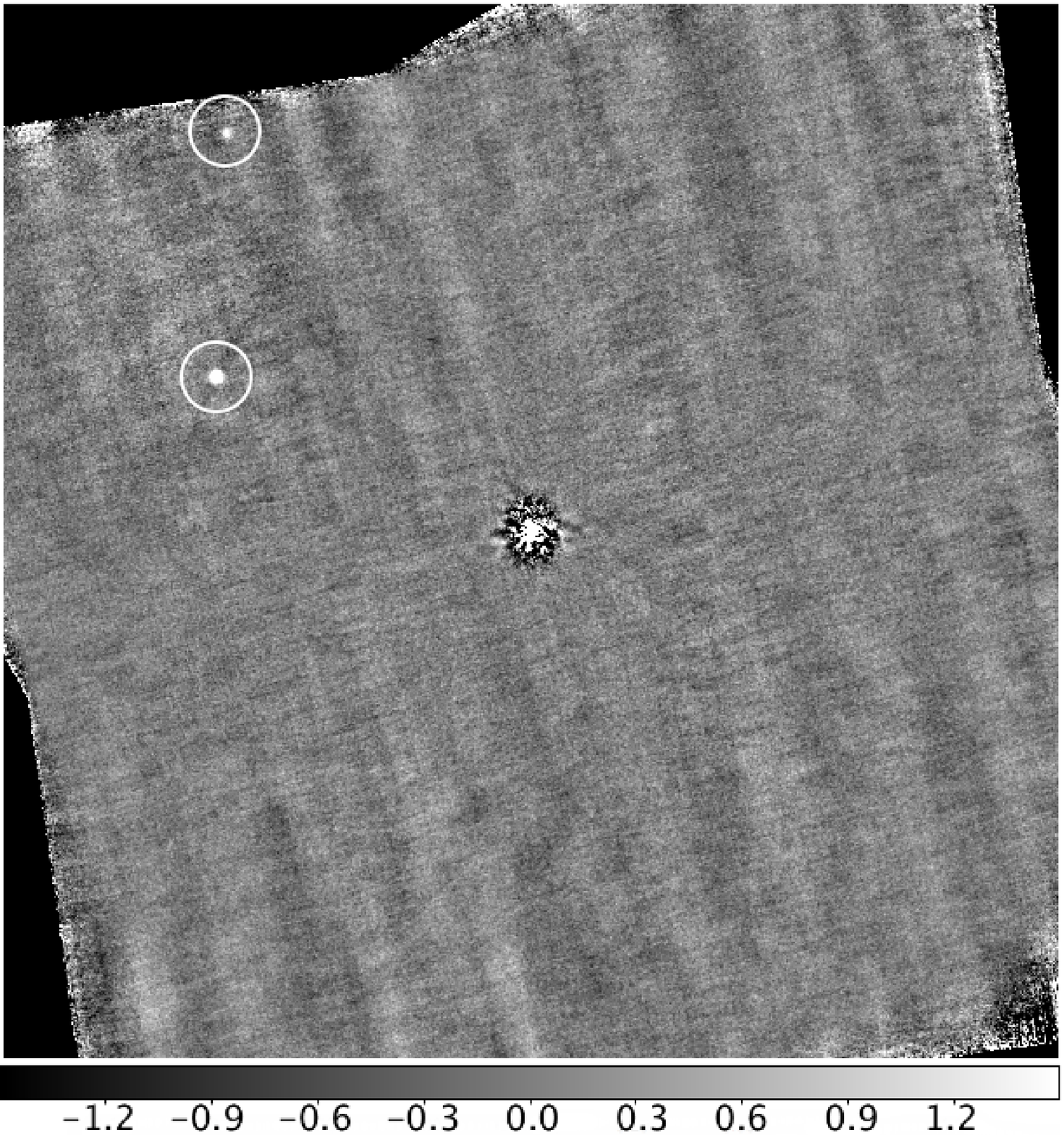}
    \label{fig:V1174Tau}}
  \subfigure[Melotte~22~SSHJ~K101]{
    \FigureFile(80mm,80mm){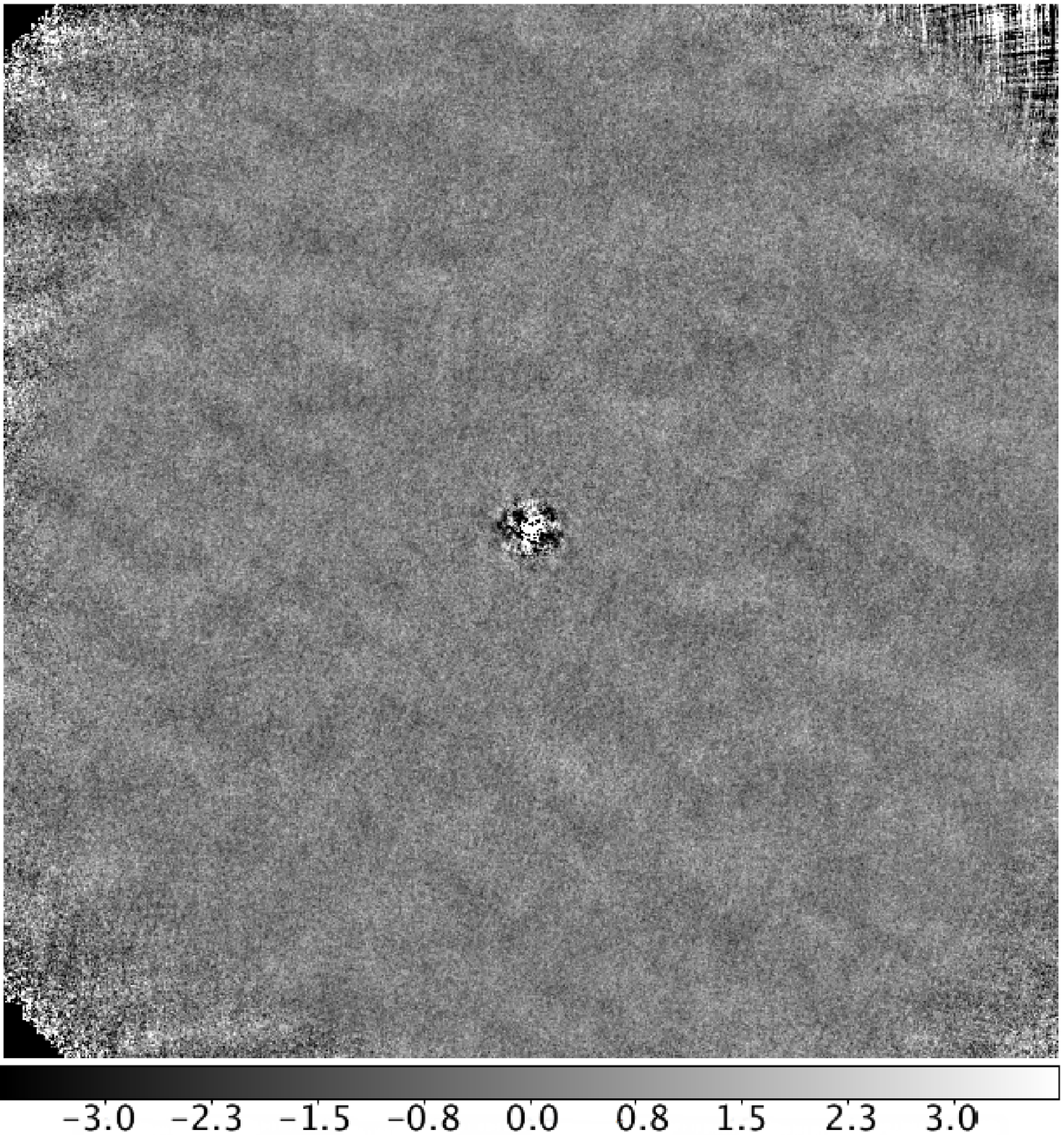}
    \label{fig:K101}}
  \end{center}
  \caption{
  {\it Continued.}
		\par\noindent
  		{\it Top left panel:} Melotte~22~SSHJ~G221.
		{\it Top right panel:} V1054~Tau.
  		{\it Lower left panel:} V1174~Tau.
  		{\it Lower right panel:} Melotte~22~SSHJ~K101.
		The unit of the color bar is ADU per each exposure time.
		}\label{fig:all results5}
\end{figure*}

\newpage
\newpage


\end{document}